\documentclass[universe,article,accept,moreauthors,pdftex]{Definitions/mdpi}

\firstpage{1} 
\pubvolume{1}
\issuenum{7}
\articlenumber{341}
\pubyear{2021}
\copyrightyear{2021}
\externaleditor{{Academic Editors: Fabio Bellini and Claudia Tomei} %MDPI: Please confirm if Academic editor should be added?
} % For journal Automation, please change Academic Editor to "Communicated by"
\datereceived{15 June 2021} 
\dateaccepted{02 August 2021} 
\datepublished{10 September 2021} 
\hreflink{https://doi.org/10.3390/universe7090341} % If needed use \linebreak
\usepackage{float}
\usepackage{xcolor}

\Title{Neutrinoless Double Beta Decay with Germanium Detectors: 10$^{26}$~yr and Beyond}
\TitleCitation{Neutrinoless Double Beta Decay with Germanium Detectors: 10$^{26}$~yr and Beyond}

\Author{Valerio D'Andrea %MDPI: Please carefully check the accuracy of names and affiliations. Changes will not be possible after proofreading.
 $^{1,2}$\orcidA{}, Natalia Di Marco $^{2,3,}$*\orcidB{}{},  Matthias Bernhard Junker $^{2}$\orcidC{}{}, Matthias Laubenstein $^{2}$\orcidD{}{}, \mbox{Carla Macolino} $^{1,2}$\orcidE{}{}, Michele Morella $^{2,3}$\orcidF{}{}, Francesco Salamida $^{1,2}$\orcidG{}{} and Chiara Vignoli $^{2}$\orcidH{}{}}

\AuthorNames{Valerio D'Andrea, Natalia Di Marco,  Matthias Bernhard Junker, Matthias Laubenstein, Carla Macolino, Michele Morella, Francesco Salamida  and Chiara Vignoli}

\AuthorCitation{D'Andrea, V.; Di Marco, N.; Junker, M.B.; Laubenstein, M.; Macolino, C.; Morella, M.; Salamida, F.; Vignoli, C.}
\address{%
$^{1}$ \quad Dipartimento di Scienze Fisiche e Chimiche, Università degli Studi dell'Aquila, 67100, L'Aquila, 
 Italy; valerio.dandrea@lngs.infn.it (V.D.); carla.macolino@univaq.it (C.M.); francesco.salamida@aquila.infn.it (F.S.) %MDPI: Newly added information, please confirm.
\\
$^{2}$ \quad INFN---Laboratori Nazionali del Gran Sasso, Assergi, 67100, L'Aquila, Italy; matthias.junker@lngs.infn.it (M.B.J.); matthias.laubenstein@lngs.infn.it (M.L.), chiara.vignoli@lngs.infn.it (C.V.)\\
$^{3}$ \quad Gran Sasso Science Institute, 67100, L'Aquila, Italy; natalia.dimarco@gssi.it (N.D.M.); michele.morella@gssi.it (M.M.)}

\corres{Correspondence: natalia.dimarco@gssi.it}

\newcommand{\ctsper}      {cts/(keV$\cdot$kg$\cdot$yr)}

\newcommand{\pIIbi  }     {{$10^{-3}$~cts/(keV$\cdot$kg$\cdot$yr)}}

\newcommand{\kgyr}        {{kg$\cdot$yr}}

\newcommand{\be}          {$\beta$}

\newcommand{\qbb}         {{$Q_{\beta\beta}$}}
\newcommand{\thalfzero}   {${T^{0\nu}_{1/2}}$}

\newcommand{\onbb}        {{$0\nu\beta\beta$}}
\newcommand{\nnbb}        {{$2\nu\beta\beta$}}

\newcommand{\gerda}       {\textsc{Gerda}}
\newcommand{\GERDA}       {\mbox{\textsc{Gerda}}}  

\newcommand{\LNGS}        {LNGS}

\newcommand{\legend}       {LEGEND} 
\newcommand{\LEGEND}       {LEGEND}
\newcommand{\majorana}    {\textsc{Majorana}}
\newcommand{\Majorana}    {{\mbox{\textsc{Majorana}}}}
\newcommand{\igex}        {\textsc{Igex}}
\newcommand{\IGEX}        {{\mbox{\textsc{Igex}}}}
\newcommand{\hdm}         {\textsc{HdM}}
\newcommand{\HDM}         {\mbox{\textsc{HdM}}}

\newcommand{\gesix}       {{$^{76}$Ge}}
\newcommand{\gess}        {{$^{76}$Ge}}
\newcommand{\geenr}       {{$^{\rm enr}$Ge}}          %$^{\rm enr}$Ge

\abstract{In the global landscape of neutrinoless double beta (\onbb)~decay search, the use of semiconductor germanium detectors provides many advantages. The excellent energy resolution, the negligible intrinsic radioactive contamination, the possibility of enriching the crystals up to 88\% in the \gess~isotope as well as the high detection efficiency, are all key ingredients for highly sensitive \onbb~decay search. The \majorana~ and \gerda~experiments successfully implemented the use of germanium (Ge) semiconductor detectors, reaching an energy resolution of 2.53~$\pm$~0.08 keV at the Q$_{\beta\beta}$ and an unprecedented low background level of $5.2 \times 10^{-4}$%MDPI: We changed the dot to multiplication sign, please confirm all in text.
~cts/(keV$\cdot$kg$\cdot$yr), respectively. In this paper, we will review the path of \onbb~decay search with Ge detectors from the original idea of E.~Fiorini et al. in 1967, to the final recent results of the \gerda~experiment  setting a limit on the half-life of \gess~ \onbb~decay at $T_{1/2} > 1.8 \times 10^{26}$~yr (90\% C.L.). We will then present the LEGEND project designed to reach a sensitivity to the half-life up to $10^{28}$~yr and beyond, opening the way to the exploration of the normal ordering region. }


\begin{document}
\section{Introduction} 
%perché ovbb panoramica tecniche  panoramica experimenti

The evidence for non-zero neutrino masses as a consequence of the neutrino oscillation discovery~\cite{PhysRevLett.81.1562,PhysRevLett.90.021802,RevModPhys.88.030502,RevModPhys.88.030501} provides, among~others, a~hint of physics beyond the Standard Model (SM). Despite many experimental efforts carried out about neutrino physics since the first pioneering experiment by Reines ad Cowan in 1956~\cite{Cowan103}, there are still open points to be clarified such as the neutrino nature, the~mass ordering and the absolute mass~scale. 

The search for neutrinoless double beta (\onbb)~decay is considered as the most promising way to prove the Majorana nature of neutrinos as well as to give an indication on the mass hierarchy and on the absolute mass scale. Moreover, the~discovery of \onbb~decay would  open the way for theories predicting the observed matter anti-matter asymmetry of the Universe being a consequence of lepton number violation through~leptogenesis. 

The important implications for particle physics and cosmology justify the experimental efforts carried out worldwide in the field of \onbb~decay searches. A~number of different experiments exploiting various technologies and searching for the transition in different isotopes exists: \gesix~\cite{science,majorana2019}, $^{82}$Se
~\cite{Azzolini:2019tta}, $^{100}$Mo~\cite{Alenkov:2019jis,Arnold:2015wpy,Armengaud:2019loe}, $^{130}$Te~\cite{Adams:2019jhp,Andringa:2015tza}, $^{136}$Xe~\cite{Anton:2019wmi,KamLAND-Zen:2016pfg,Martin-Albo:2015rhw} and~others.   

In this paper, we will review the story of the \onbb~decay search of \gesix\ with germanium semiconductor detectors. The~paper is organized as follows. In~the second section, a review of the main aspects of the \onbb\ theory will be presented. In~the third section, we will discuss the choice of the \gesix\ isotope, highlighting the main advantages and disadvantages with respect to other techniques and isotopes. In~the fourth section, we will report about the historical path of \onbb~decay search with germanium detectors from the original idea proposed by E.~Fiorini et al. in 1967~\cite{fiorini1967} to the  experiments inaugurating the modern era of highly sensitive, low background searches. In~the fifth and sixth section, we will summarize the main characteristics, performance and results of the contemporary \gerda\ and \majorana\ experiments currently leading the field with the best sensitivity, the~lowest background level and the best resolution among all the other \onbb~decay experiments. Finally, in~the seventh section, we will present the \legend\ project conceived to extend the sensitivity up to $10^{28}$~yr to fully cover the inverted hierarchy~region.

\section{Neutrinoless Double Beta~Decay}
\begin{figure}[H]
      \begin{tabular}{cc}
    \includegraphics[width=0.33\textwidth]{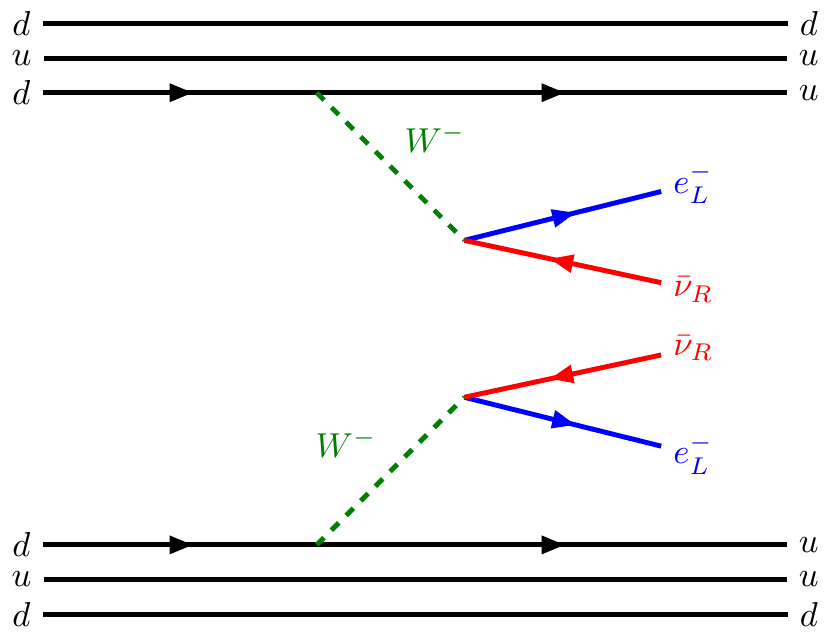}
    &
    \includegraphics[width=0.33\textwidth]{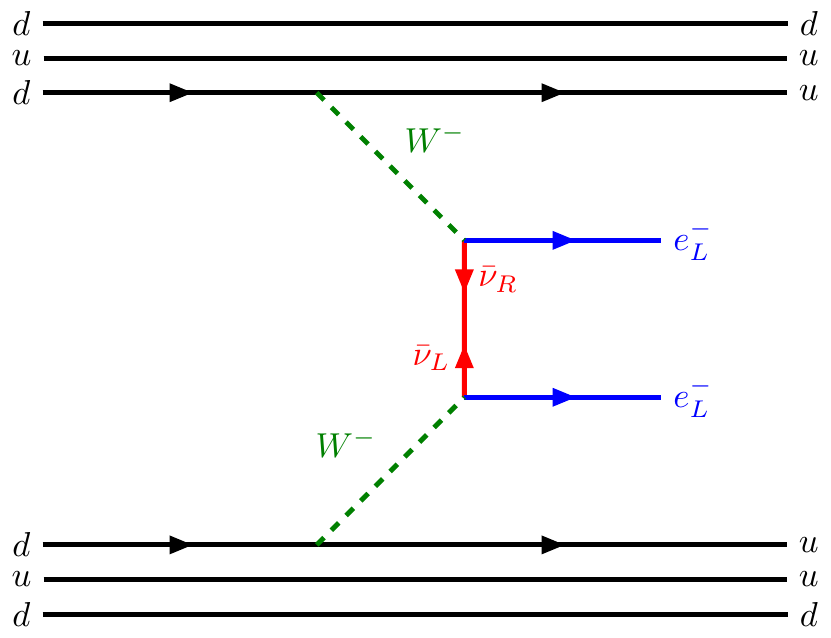}
    \end{tabular}
    \caption{On the \textbf{left}, the~diagram of the standard \nnbb\ decay with the emission of 2 anti-neutrinos. On~the \textbf{right}, the~\onbb\ process due to the exchange of a light Majorana neutrino  with lepton number violation.}
    \label{fig:bb_decay}
\end{figure}

The double beta decay (\nnbb), first proposed by M. Goeppert-Mayer in 1935~\cite{2nbb-GoeppertMayer1935}, is a second-order process generated in the perturbative expansion of weak interactions in  the SM with an extremely long lifetime corresponding to a transition from a nucleus $(A,Z)$ to $(A,Z+2)$. Candidate isotopes that can decay through the double beta decay are even--even nuclei and lighter than the odd--odd $(A,Z+1)$ nucleus, for~which the single $\beta$ decay is forbidden or strongly suppressed because of a large change of spin. In~\nnbb~decay, starting from the initial nuclear state, one nucleon type (proton or neutron) decays into another one emitting a lepton--anti-lepton pair through a virtual transition, and a second decay occurs producing a further pair of lepton and anti-lepton. The~SM, therefore, predicts the emission in the final state of two electrons and two anti-neutrinos:
\begin{equation*}
  (Z,A) \rightarrow (Z+2, A) + 2e +2\bar{\nu} + Q_{\beta\beta}
\end{equation*}
where \qbb\ is the energy released in the decay. The~Feynman diagram of this process is shown in the left panel of Figure~\ref{fig:bb_decay}. Among~the 35 naturally occurring isotopes that can decay through \nnbb, Table~\ref{tab.2nubb} reports the results for the measurement of a few isotopes together with the main characteristics of the~process. 

The \onbb~decay process, proposed by W.H. Furry in 1939~\cite{onbb-Furry1939}, is a SM-forbidden transition in which the final state consists of only two electrons:
\begin{equation*}
  (Z,A) \rightarrow (Z+2, A) + 2e + Q_{\beta\beta}
\end{equation*}

This decay clearly violates the leptonic number conservation ($\Delta L=2$). The~relevant diagram, due to the exchange of light Majorana neutrino,  is shown in Figure~\ref{fig:bb_decay} on the right panel. 
There are other possible mediators of this decay, e.g.,~right-handed weak currents, supersymmetric particles and massive neutrinos. Regardless of the underlying mechanisms, the~observation of the \onbb~decay would prove that the lepton number conservation is not an exact symmetry of nature and that neutrinos have a Majorana mass~component.  
%Neutrinoless double beta decay-lifetime-phase space factors- nuclear matrix elements\\

\begin{specialtable}[H] 
\caption{A list of isotopes mostly used in \nnbb\ decay experiments. Some of the key features are reported. Half-life values are from%MDPI: Please make sure that permission has been obtained and there is no copyright issue. please check all in text
~\cite{barabash2020precise}.\label{tab.2nubb}}
\setlength{\tabcolsep}{6.9mm}
\begin{tabular}{cccc}
\toprule
\textbf{Isotope}	& \textbf{Natural} & \textbf{Q\boldmath{$_{\beta\beta}$}} & \textbf{T\boldmath{$^{2\nu}_{1/2}$}}\\
 & \textbf{abundance} (\boldmath{$\%$}) & \textbf{(MeV)} & \textbf{(yr)}\\ 
\midrule
$^{48}Ca$  & 0.187 & 4.263 & $5.3^{+1.2}_{-0.8} \times 10^{19}$ \\
$^{76}Ge$  & 7.8   & 2.039 & $(1.88\pm 0.08) \times 10^{21}$ \\
$^{82}Se$  & 9.2   & 2.998 & $0.87^{+0.02}_{-0.01} \times 10^{20}$ \\
$^{96}Zr$  & 2.8   & 3.348 & $(2.3 \pm 0.2) \times 10^{19}$ \\
$^{100}Mo$ & 9.6.  & 3.035 & $7.06^{+0.15}_{-0.13} \times 10^{18}$ \\
$^{116}Cd$ & 7.6   & 2.813 & $(2.69 \pm 0.09)\times10^{19}$ \\
$^{130}Te$ & 34.08 & 2.527 & $(7.91 \pm 0.21) \times 10^{20}$ \\
$^{136}Xe$ & 8.9   & 2.459 & $(2.18 \pm 0.05)\times10^{21}$ \\
$^{150}Nd$ & 5.6   & 3.371 & $(9.34 \pm 0.65)\times10^{18}$ \\
\bottomrule
\end{tabular}
\end{specialtable}
%\subsection{Lifetime of neutrinoless double beta decay}

In the low energy limit, the interaction of neutrinos can be described by the current--current four-fermion interactions. In~this approximation, the half-life of the \onbb\ decays can be derived as described in~\cite{10.1143/PTP.66.1765}:
\begin{equation}
\left(T_{1/2}^{0\nu}\right)^{-1} = {G_{0\nu}} \times {\left| M_{0\nu}\right|^{2}} \times { \left(\frac{m_{\beta\beta}}{m_{e}}\right)^2 }  
\end{equation}
where $G_{0\nu}$ is the phase space factor (PSF), $M_{0\nu}$ is the nuclear matrix element (NME) and $m_{\beta\beta}$ is the effective Majorana mass defined by:
\begin{equation}
    m_{\beta\beta} = \sum_{i=1}^3 U_{ei}^2 m_i
 \end{equation}
where $U$ is the PNMS mixing matrix and $m_i$ are the neutrino mass eigenvalues. It follows that the decay rate is proportional to $m^2_{\beta\beta}$ and so the neutrino mass can be estimated from the measurement of the half-life of the \onbb~decay.
%\subsection{Nuclear Matrix Elements}

\begin{figure}[H]
    \includegraphics[width=0.7\textwidth]{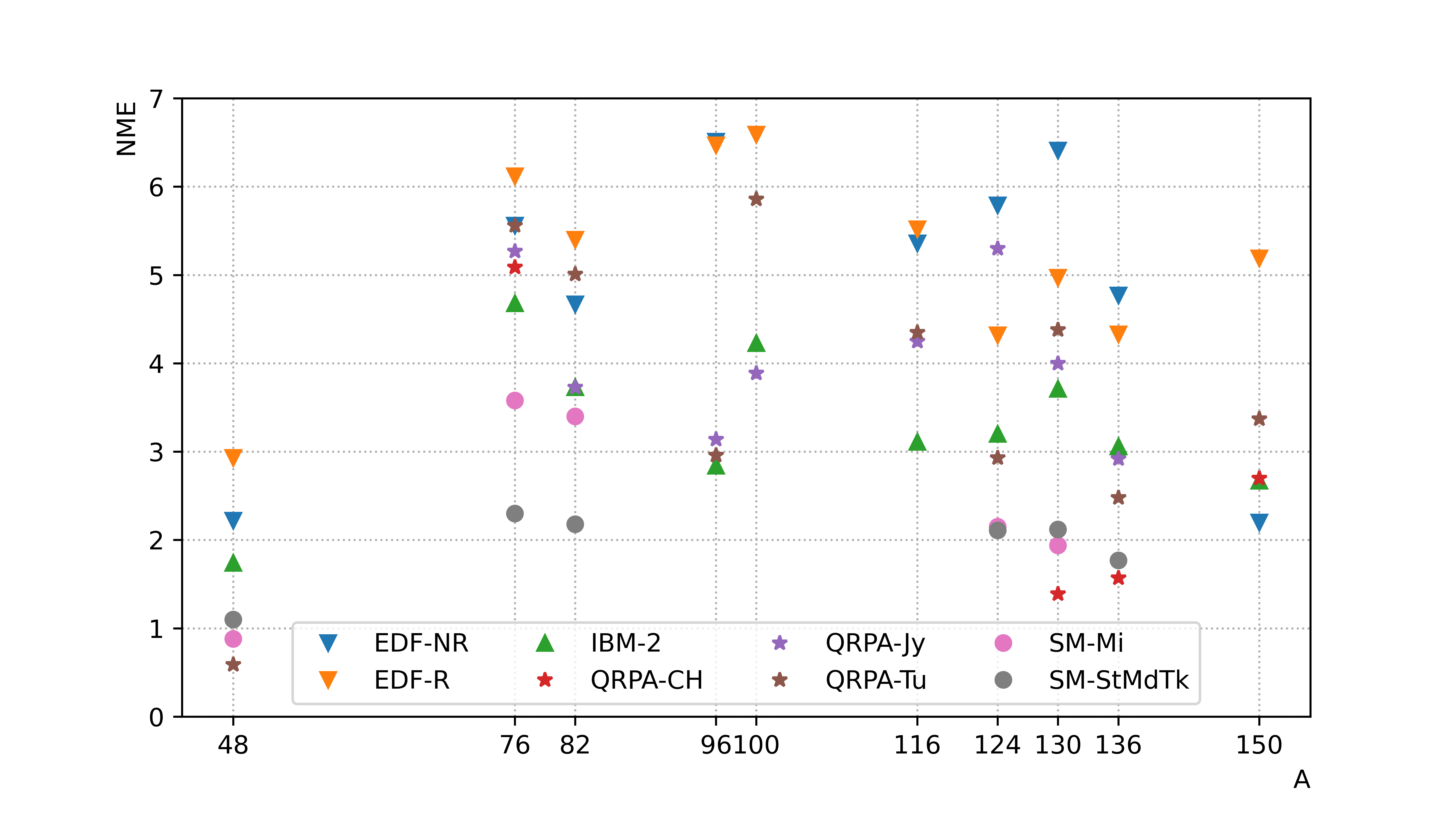}
        \caption{Comparison of NME calculations with an unquenched $g_A = 1.27$ for different isotopes and with various models. Abbreviations: EDF, energy-density functional; IBM, interacting boson model; NME, nuclear matrix element; QRPA, quasi-particle
random-phase approximation; SM, Standard Model. The~reader should refer to%MDPI: Please make sure that permission has been obtained and there is no copyright issue.
~\cite{Dolinski:2019nrj} for further details.}
    \label{fig:nme}
\end{figure}

The main source of uncertainty in the expected value of the \onbb~decay half-life originates from the NME calculations. They involve the hadronic current of the weak Hamiltonian and depend on the nuclear-structure of the parent and daughter nuclei. They are usually difficult to calculate even in the case of single $\beta$~decay.  Different models are available to calculate the NMEs, e.g., the Interactive Shell Model (ISM)~\cite{MENENDEZ2009139}, the~Quasi-particle Random Phase Approximation (QRPA)~\cite{Simkovic:2013qiy} and the Interactive Boson Model (IBM-2)~\cite{PhysRevC.91.034304}.
A comparison between the values of the NMEs calculated with the different models is illustrated in Figure~\ref{fig:nme}. Despite the fact that the uncertainty on the single calculation is of the order of 20\%, the differences between the different models are even larger. The~overall estimate of the NME calculation is therefore affected by a large systematic~uncertainty.

Parametrizing the NME as in~\cite{PhysRevC.60.055502}, we can write:
\begin{equation}\label{eq.ga}
M_{0\nu} = g^2_A\left( M^{0\nu}_{GT} - \left( \frac{g_V}{g_A}\right)^2 M_F^{0\nu} + M_T^{0\nu}\right)
\end{equation}
where $g_V$ and $g_A$ are the vector and axial coupling constants of the nucleon, $M_{GT}^{0\nu}$ is the Gamow--Teller operator matrix element between the initial and the final states, $M_F^{0\nu}$  is the Fermi contribution and $M_T^{0\nu}$ is the tensor operator matrix element. \mbox{Equation~(\ref{eq.ga})} emphasizes the role of $g_A$ being  $T_{1/2}^{0\nu}$ proportional to $g_A^{-4}$. The~value of $g_A$ can be used as an adjustment to reconcile observations with calculations. In~the hypothesis of quenching, a reduction of $g_A$ is assumed to reproduce the observable quantities of $\beta$ and \nnbb~decays~\cite{PhysRevD.97.035005}. There are different possible origins for  quenching, e.g.,~many-body currents, intrinsic shortcomings of the nuclear many-body models, or~nuclear medium effects~\cite{Dolinski:2019nrj}. Recent studies point towards quenching not being larger than 20–30\% in processes with high momentum transfer such as the \onbb~decay. This reduction would result in an increase of $T_{1/2}^{0\nu}$ by a factor of two or~three.

\section{The Choice of the \gesix\ Isotope}
\begin{figure}[H]
      \includegraphics[width=0.6\textwidth]{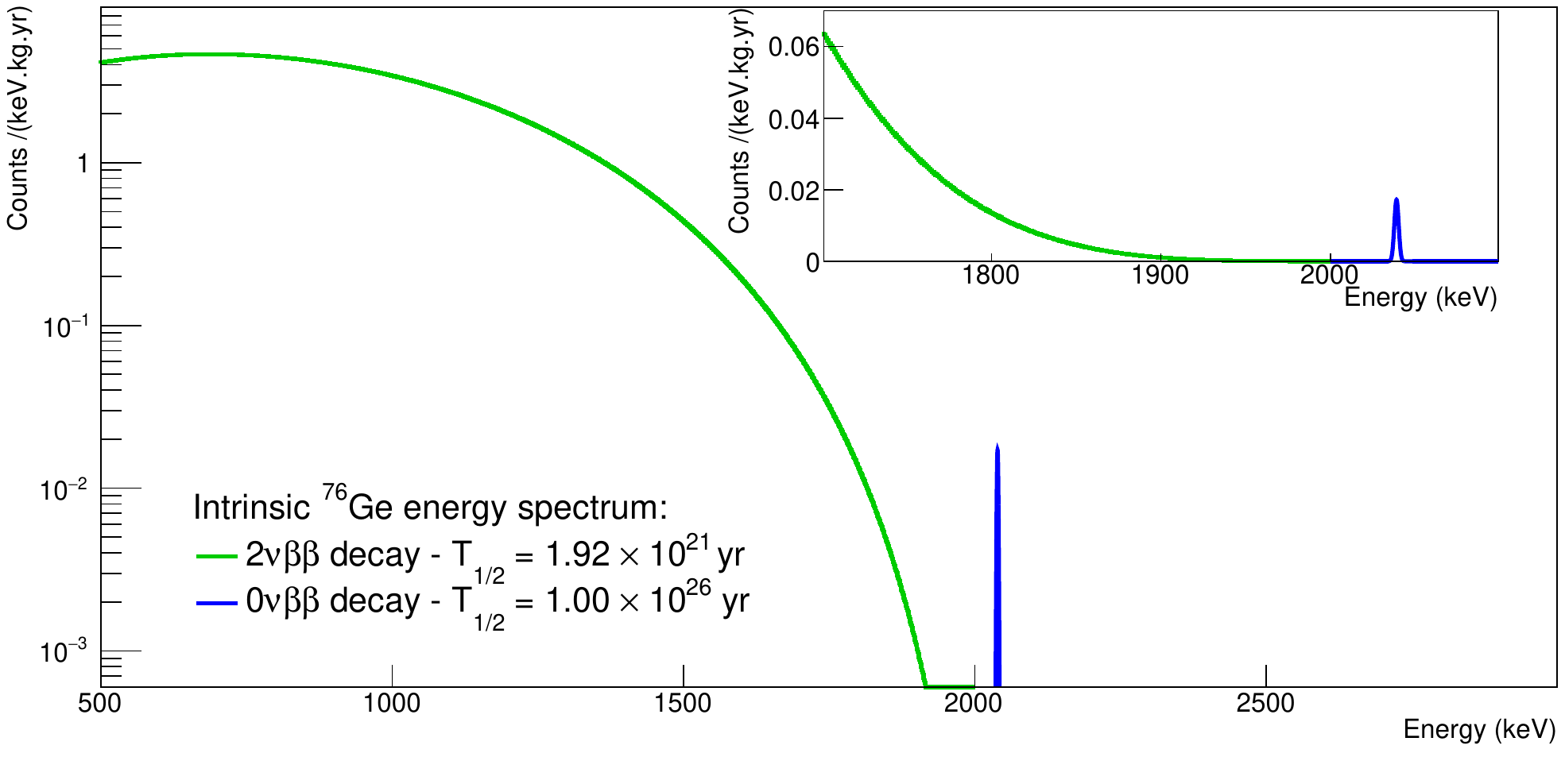}
        \caption{{Schematic view of}  the expected \onbb\ and \nnbb~decay spectra. Credits to \gerda\ \mbox{Collaboration}.}
    \label{fig:0bb_exp}
\end{figure}

From the experimental point of view, the~search for \onbb~decay consists of the detection of the two emitted electrons. As the recoil of the nucleus negligible, the~sum of energies of the two electrons corresponds to the Q-value of the process. The~signature of the \onbb~decay is therefore a mono-energetic peak centered at \qbb\ (see Figure~\ref{fig:0bb_exp}).
Since the value of \qbb\ is well known, the~search for a \onbb~decay signal can be performed in a very narrow region of interest (ROI). The~number of signal events expected in the ROI is given by:
\begin{equation}\label{eq.sens1}
N^{0\nu} = \text{ln(2)}~\frac{N_A}{m_A}\left(\frac{M~T~\varepsilon~a}{T_{1/2}^{0\nu}} \right) 
\end{equation}
where $N_A$ is the Avogadro constant, $m_A$ and $a$ are the molar mass and the isotopic abundance of the double beta ($\beta\beta$) emitter, respectively, $\varepsilon$ is the detection efficiency, $M$ is the detector mass and $T$ is the measurement time of the experiment.
On the other hand, the number of background events in the ROI is given by:
\begin{equation}\label{eq.sens2}
N_{b} = \text{BI}~M~T~\Delta E 
\end{equation}
where $\Delta E$ is the ROI width, proportional to the energy resolution of the detector, and~BI is the background index, normalized to the width of the ROI, mass, energy and time and expressed in units of keV$^{-1}$kg$^{-1}$year$^{-1}$.
The sensitivity of the experiment can be obtained requiring that the \onbb~decay signal exceeds the standard deviation of the total detector counts $N^{0\nu} \geq n_{\sigma}~\sqrt{N^{0\nu} + N_b}$ where $n_{\sigma}$ is the confidence level expressed in units of $\sigma$s of the Poisson distribution. Combining Equations~(\ref{eq.sens1}) and ~(\ref{eq.sens2}), the sensitivity can be derived as:
\begin{equation}\label{eq.sens3}
S^{0\nu} = \text{ln(2)}~\frac{N_A}{m_A}\left(\frac{\varepsilon~a}{n_\sigma} \right) \sqrt{\frac{M~T}{\text{BI}~\Delta E}}
\end{equation}

This formula clearly shows the role of the different experimental parameters and highlights the different characteristics that have to be chosen to improve the \onbb~decay discovery probability. In~particular, the ideal experiment should have an isotope with high isotopic abundance $a$, a~high detection efficiency $\varepsilon$, the~possibility to deploy a large amount of mass, a~low background and a good energy~resolution.

Among $\beta\beta$-emitters, $^{130}$Te has the highest isotopic abundance and a tellurium-based experiment does not necessarily require enrichment. Tellurium dioxide crystals ($^{nat}$TeO$_2$), as the ones employed in CUORE~\cite{Adams:2019jhp}, are used as cryogenic calorimeters detecting the small temperature rise induced by the energy deposition of a charged particle. As~in all  experiments where the source of  \onbb-decay events coincides with the detector, the~detection efficiency is maximized.
Although reaching a very good energy resolution (\hbox{$\sim$0.3\% FWHM at \qbb,} second only to germanium diodes), phonon detection only does not allow to perform particle identification to reject background events. This leads to a higher background index and prevents entering the background-free regime. Moreover, the  \qbb~of~$^{130}$Te (see Table~\ref{tab.2nubb}) is lower than the $^{208}$Tl gamma line at 2.6 MeV, which therefore contributes to the background around \qbb. The~solution to overcome this limitation is the use of scintillating bolometers employing different isotopes, namely Zn$^{82}$Se for CUPID-0~\cite{Azzolini:2019tta} and Li$^{100}$MoO$_4$ for CUPID-Mo~\cite{Armengaud:2019loe}. %, which are the two possibilities considered for next generation ton-scale experiment CUPID. 
Both of these isotopes have a \qbb~ around 3 MeV, above the $^{208}$Tl line and most of the other natural radioactive background sources, but~very low natural isotopic abundance (in this case, enrichment is required). Li$^{100}$MoO$_4$ crystals will be the ones implemented in the final version of the next generation ton-scale experiment CUPID~\cite{CUPID:2019imh}. This is mainly due to the poor radiopurity and energy resolution of Zn$^{82}$Se produced, even though $^{100}$Mo $2\nu\beta\beta$ decay half-life (the fastest among the ones in Table~\ref{tab.2nubb}) could be an irreducible source of background around \qbb~ if energy and time resolution requirements will not be~satisfied.

Although having a \qbb~below $^{208}$Tl, $^{136}$Xe is an interesting candidate isotope as it is gaseous, easy to enrich and can be used to build a ton-scale experiment. Also  in this case, the~detector itself is the source of \onbb-decay, and different technologies can be implemented: a single phase Time Projection Chamber (TPC) using Liquid Xenon (LXe), such as EXO-200~\cite{EXO200}, a~high-pressure TPC with Gaseous Xenon (GXe), such as NEXT~\cite{NEW}, or a Xe-loaded liquid scintillator, such as  KamLAND-Zen~\cite{Gando:2020cxo}. The~achievable energy resolution is lower with respect to germanium or bolometric experiments and it ranges roughly from 1\% to 10\% FWHM among different experiments. LXe-TPCs measure both the scintillation light (starting time for the TPC) and ionization charges and have a better self-shielding against external background sources (due to the higher density of the target). Furthermore, some topological information to distinguish single site from multi site events was used in EXO-200~\cite{EXO200}.  GXe-TPCs allow for real topological reconstruction: because of the lower density of the detector, a~\onbb~event would look like two separate tracks (originating from the same point, the~decaying nucleus) ending with Bragg peaks, corresponding to the stopping of the two emitted electrons. In~this way, it is possible to reject nearly any possible background event except  $2\nu\beta\beta$. Xe-based experiments did not reach a background-free regime up to now; thus, the \onbb~decay  (and also any other physics) search heavily depends on the background model of the~experiment. 

Germanium detectors are particularly suited for \onbb~decay search, presenting  indeed several advantages. As the search for \onbb~decay is based on the detection of a signal peak over the background, the~excellent  energy resolution of semiconductor detectors results in a great advantage. High Purity Germanium (HPGe) detectors have the best energy resolution with respect to any other competitive technique with a full width at half maximum (FWHM) better than 0.1$\%$ at the $Q$-value of the \onbb~decay of \gesix\ (\qbb ~= 2039~keV) (see Sections~\ref{sec:gerda} and~\ref{sec:majorana}). This feature allows to identify the  $\gamma$ peaks of the various background sources as well as to isolate the tail of the \nnbb~decay spectrum (see \mbox{Figure~\ref{fig:0bb_exp}}). Germanium-based experiments feature a high detection efficiency since the detector is also the source of the $\beta\beta$-decay.  Despite the relatively low natural isotopic abundance of the germanium $\beta\beta$-emitter \gesix\ (7.8\%), modern experiments make use of crystals enriched in \gess\ (\geenr) up to 88\%, thus significantly reducing the number of detectors needed to reach a given \gesix\ content.
HPGe detectors also guarantee a low background level since they have an extremely high intrinsic radio-purity (no measurable U or Th contamination) and can be realized with particular electrode geometry, allowing to take advantage of the pulse shape discrimination (PSD) analysis to actively reduce the background (see \mbox{Sections~\ref{sec:gerda} and~\ref{sec:majorana}}).

On the other hand, the use of germanium presents a few disadvantages. The~\qbb\ of \gesix\ is low with respect to other  commonly used $\beta\beta$ isotopes. In~particular, the \qbb\ is below the dominant $^{208}$Tl line of 2615~keV, thus causing the ROI suffering of the relative Compton continuum. The~enrichment process is a rather expensive procedure even if the cost has been decreasing through the years. Although~taking advantage from a high value of $M_{0\nu}$, \gesix\ has the lowest value of the space phase factor ($\sim$2.3~$\times$~10$^{-15}$~\cite{Kotila:2012zza,Stoica:2013lka}) with respect to all other isotopes. This implies the need to reach a longer \thalfzero\ to probe a given $m_{\beta\beta}$ value.

Modern experiments using HPGe detectors have already faced the listed issues and are currently leading the field with the best limit and the best sensitivity on the \onbb\ decay half-life with the \gerda\ experiment (see Section~\ref{sec:gerda}), thus approaching the exploration of the inverted ordering region.
PSD techniques and additional active veto systems have been successfully exploited to get rid of $\gamma$ background and other residual contamination. These factors, together with a careful material selection, led \gerda\ to achieve an unprecedented low background level thus, reaching the so-called background-free %MDPI: Is the italics necessary? if not, please remove it.
regime (see Section~\ref{sec:gerda}).

\section{Neutrinoless Double Beta Decay Search with Ge~Detectors}\label{sec:historical_Ge_review}

The observation of the \be\be-decay can be performed through geochemical, radiochemical and direct techniques. In~the geochemical method~\cite{primakoff1965}, the abundance of the daughter isotope is determined in  minerals of known age containing the parent nucleus.  In~the radiochemical method, the~abundance of the final nucleus is measured after several years in a well-prepared artificial
sample of the parent one. In~both methods, however, \onbb\ and \nnbb~decays cannot be distinguished, while in~direct experiments, the~measurement of the sum energy of the two emitted electrons provides a real-time detection of the \be\be-decay.

The search for \onbb~decay with Ge detectors was firstly proposed in 1967 by the Milano Group~\cite{fiorini1967} using a Ge(Li) detector~\cite{fiorini1970,fiorini1973}. The~experiment was located in the Mont Blanc tunnel ($\sim$4200~m.w.e.) in order to reduce cosmic-ray background. The~90~g Ge detector was surrounded by a plastic scintillator veto and shielded with  10~cm of low background lead, a~thin
cadmium neutron absorbing layer and a 10~cm-thick box of resin impregnated wood as a neutron moderator. The~outer shield was 10~cm-thick ordinary lead~\cite{Avignone:2019phg}.
In the following years, the Milano group upgraded the experiment with two true coaxial Ge(Li) detectors and made several improvements regarding the use of low-background construction materials for the cryostat and a new shielding configuration~\cite{bellotti1984,bellotti1986}. No evidence for \onbb~decay of \gesix\ was found, and a lower limit on the half-life of $3.3 \times 10^{23}$~yr (68\% C.L.) was set in 1986~\cite{bellotti1986}.

In the 1970s and 1980s, several research groups started to search for the \onbb~decay of \gess~with HPGe detectors, making strong efforts to reduce the background through the use of passive shields, by~placing the experiments in underground facilities~\cite{caltech1984,gdk1984,gotthard1989,gotthard1992} and by using an active NaI veto~\cite{avignone1983,avignone1985,avignone1986,caldwell1987,caldwell1991}. The~strongest limit on the \onbb~decay half-life reported by a natural Ge experiment was $1.2 \times 10^{24}$~yr (UCSB/LBL~\cite{caldwell1991}).
%The PNL/USC collaboration  used an intrinsic Ge detector in a low-background cryostat inside of a NaI(Tl) anticoincidence shield.
%At Caltech for the first time a coaxial  serves as a source (natural abundance of \gess~is 7.76\%) as well as the detector. 
%In the present experiment a specially constructed high-purity Ge detector was used in an underground site. The detector consists of a 208-cm active volume Ge in a J-shaped cryostat assembly made of materials selected for low intrinsic radioactivity. The detector was installed in a salt mine near Windsor, Ontario, at a depth of about 330~m GDK (Ontario) ~\cite{}.
%UCSB-LBL~\cite{caldwell1987,caldwell1991}

For the first time, the ITEP-Yerevan experiment~\cite{vasenko1990} employed  Ge(Li) detectors isotopically enriched in \gess~in~order to concentrate the active source mass in small detectors and set a limit of \thalfzero\ $\geq 2.0 \times 10^{24}$~yr  in 1989.% This gave the basis for the following step in the \onbb~decay search with \gess.

Following this idea, in~the 1990s, two  collaborations produced the first \geenr~HPGe detectors: the Heidelberg--Moscow (\hdm) experiment~\cite{hdm1992, hdm1995, hdm1997, hdm2001, hdm2005} and the International Germanium Experiment (\igex)~\cite{igex1998, igex2002}. The~\hdm~experiment operated 16.9~kg of \geenr~with a \gess~abundance of 86\% at the LNGS. The~\igex~collaboration produced six \geenr~detectors that were operated in a low-background cryostat with archaeological lead shielding in different underground facilities. Both \hdm~and \igex~experiments applied pulse shape discrimination (PSD) analysis to further reduce the background. A~lower limit on the half-life of the decay of $1.9\times 10^{25}$~yr (90\% C.L.) was found by the \hdm\ collaboration~\cite{hdm2001}.

In 2001, after~the publication of the final results of \hdm\, part of the collaboration published a claim on  the observation of the \onbb~decay of  $^{76}$Ge~\cite{klapdor2001}, reporting a half-life of $T^{0\nu}_{1/2} = 1.19^{+0.37}_{-0.23} \times 10^{25}$~yr~\cite{klapdor2004}. Later, pulse shape discrimination was used to strengthen the claim~\cite{klapdor2006}.
This claim arose a number of replies and was strongly criticized by many physicists~\cite{schwingenheuer2013}. The~situation was clarified only a few years later by the results from the first phase of the \gerda\ experiment~\cite{prl} that strongly disfavored the~observation.

The \gerda~\cite{prl, nature, PRL2018, science, gerdafinal} and \majorana~\cite{majorana2018,majorana2019} experiments continued the search for the \gess\ \onbb\ decay with Ge detectors in the last decade, reaching a sensitivity larger than $10^{26}$~yr. Details on these experiments are presented in Sections~\ref{sec:gerda} and \ref{sec:majorana}.

\begin{figure}[H]
\includegraphics[width=0.65\textwidth]{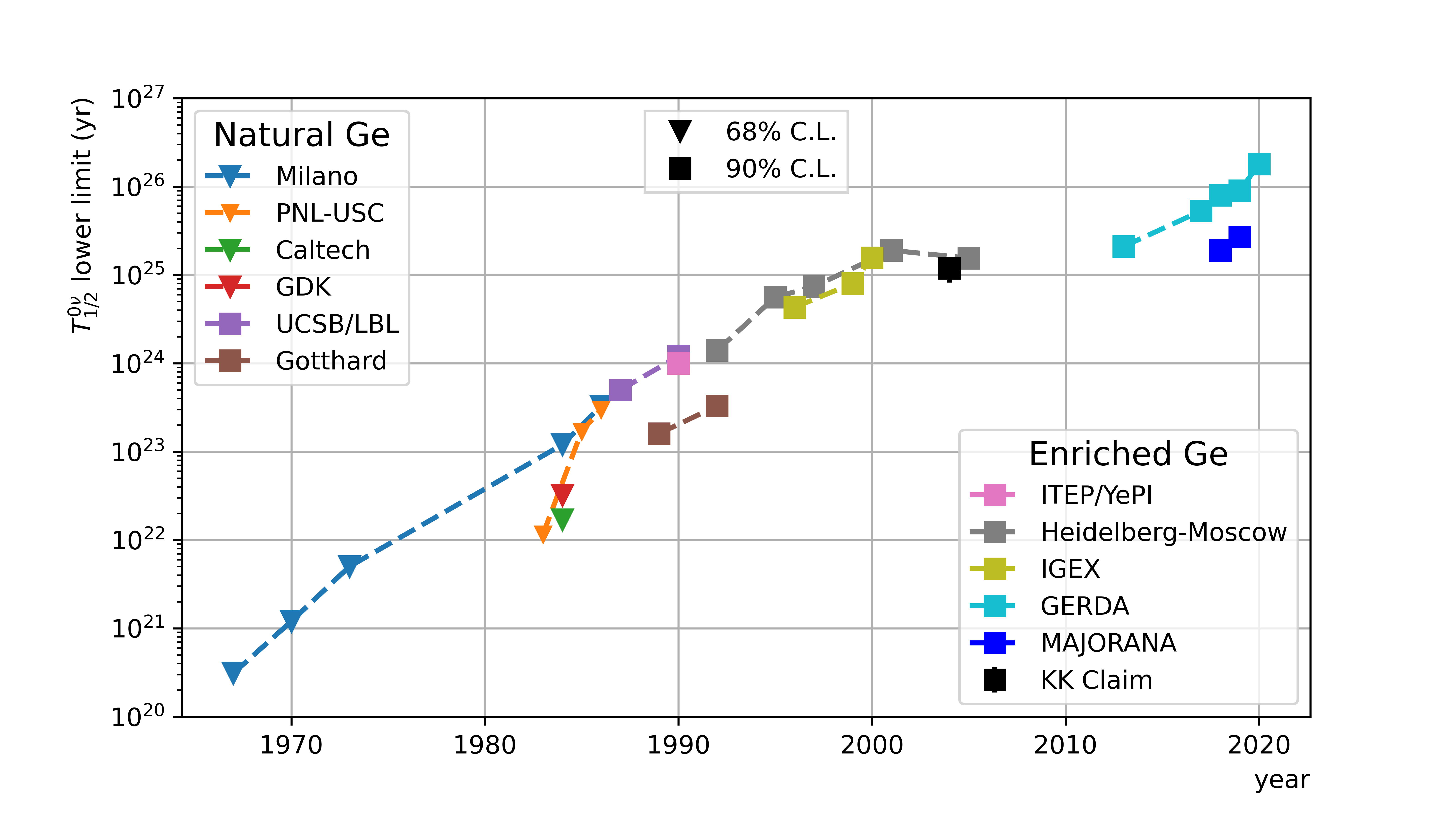}
\includegraphics[width=0.65\textwidth]{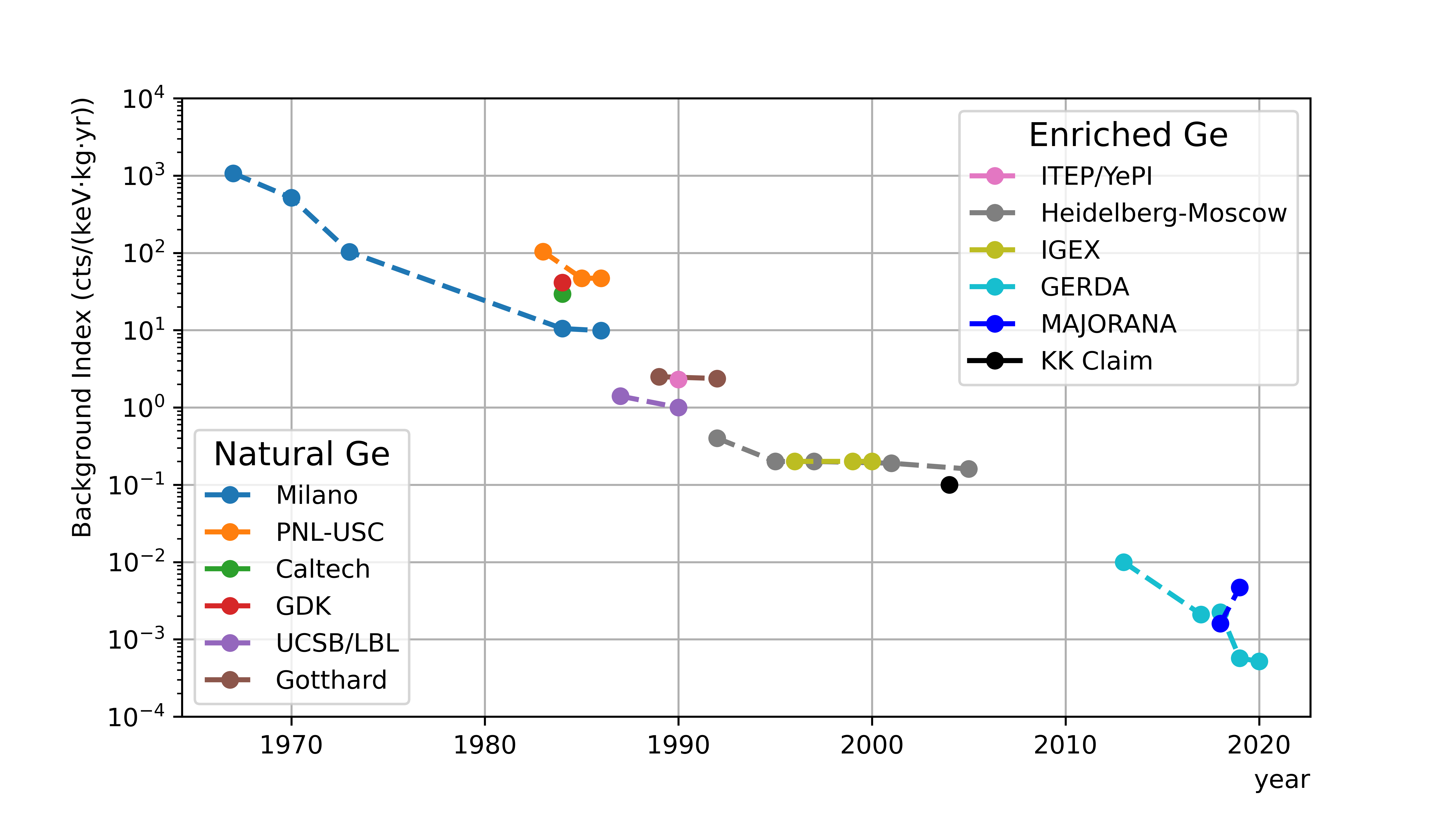}
\caption{\label{Fig:history} Evolution of the lower limit on the half-life (\textbf{upper} panel) and background index (\textbf{bottom} panel) as a function of time for the experiments searching for \gesix\ \onbb~decay. References are in the text (see Section~\ref{sec:historical_Ge_review}).} 
\end{figure}

The evolution of the published results on the lower limit of the half-life of \gess\ \onbb\ decay is depicted in Figure~\ref{Fig:history}, from~the first result of the Milano group~\cite{fiorini1967} until the final result of the \gerda~experiment~\cite{gerdafinal}.
The intense experimental program and the technology progress allowed  tightening the limit of six orders of magnitude in about 50 years. This goal has been achieved thanks to the excellent performance of the Ge detectors and the possibility of operating an increasing \gess~mass, together with the impressive improvement in the background reduction techniques (e.g., strong material selection, introduction of active vetoes and efficient pulse shape discrimination analysis). Figure~\ref{Fig:history} (bottom) reports the evolution in time of the background index of germanium  experiments: the BI was progressively decreasing as the sensitivity increased until~reaching the background-free regime with the \gerda~experiment~\cite{nature}.

The search for \onbb~decay in \gess\ will be continued in the following years by the \legend\ project~\cite{legend}: with a staged approach, it aims to reach a sensitivity to the \onbb~decay  half-life up to $10^{28}$~yr. The~experimental program of \legend\ is described in Section~\ref{sec:legend}.

\section{The GERDA~Experiment} % Natalia
\label{sec:gerda}
The experimental apparatus of the GERmanium Detector Array (\GERDA) experiment is installed in hall A of the Gran Sasso National Laboratory (\LNGS). The~experiment was taking data from 2011 to 2019 through different phases. The~facility is currently being upgraded to host the first phase of the \LEGEND\ project, called \LEGEND-200 (see \mbox{Section~\ref{sec:legend}}).    

The core of the \GERDA\ experiment was made of HPGe detectors isotopically enriched in \gess\ up to $\sim$87\%~\cite{gerdatech,gerdaupgrade}. Following a suggestion from~\cite{Heusser:1995wd}, the~detectors were directly immersed into liquid Argon (LAr), which acts both as a shield against the external radioactivity and as a cooling medium. In~the different phases of the experiment, the germanium mass was progressively increased using different types of~detectors. 

\begin{figure}[H]
\includegraphics[width=12 cm]{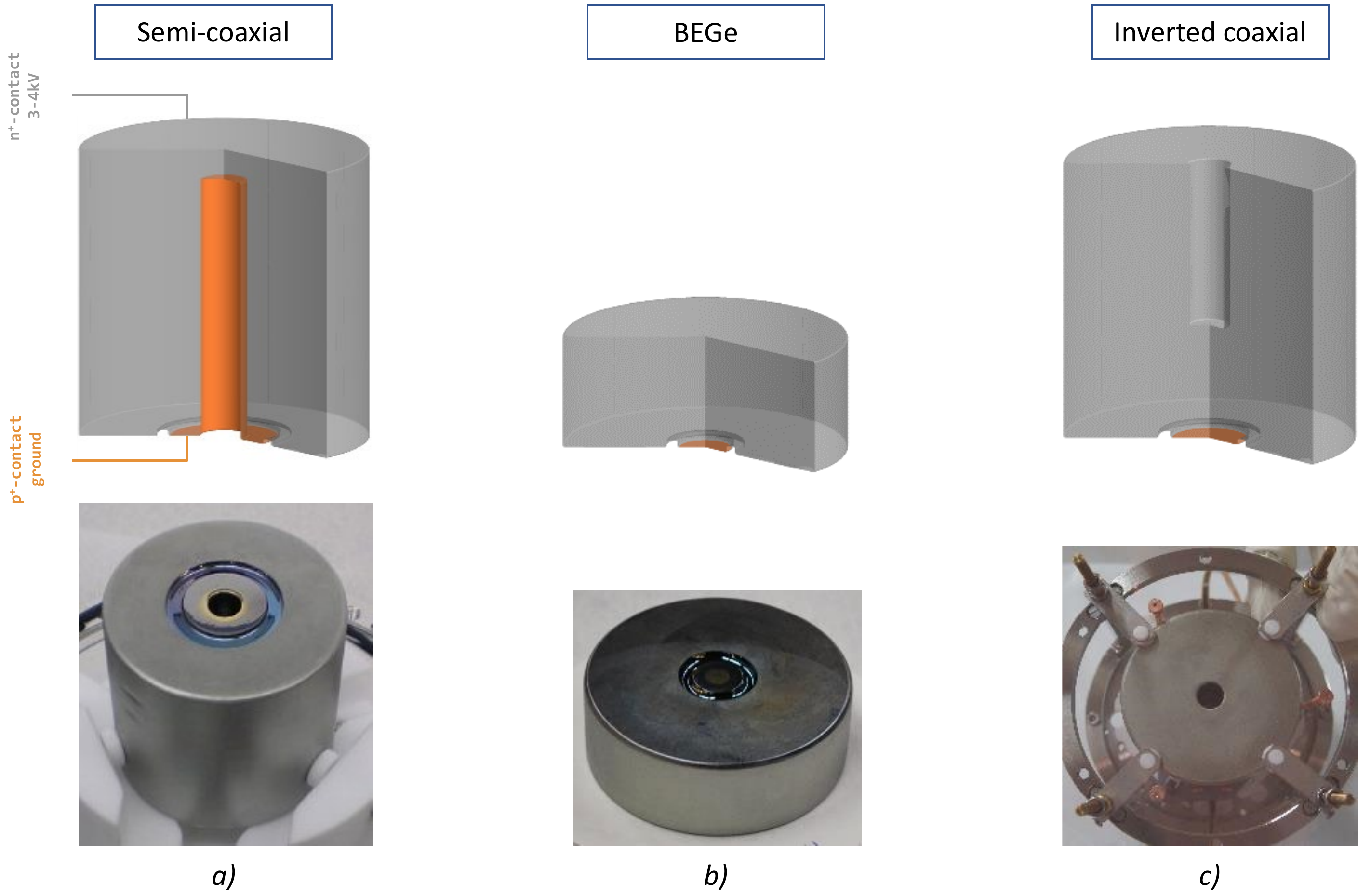}
\caption{Sketch (\textbf{top}) and picture (\textbf{bottom}) of semi-coaxial (\textbf{a}), BEGe (\textbf{b}) and inverted coaxial (\textbf{c}) detectors employed in the various phases of the \GERDA\ experiment. \label{GERDA_det}}
\end{figure}

In \GERDA\ Phase I, lasting from November 2011 to September 2013 and collecting an exposure of 23.5~\kgyr, eight enriched semi-coaxial Ge detectors (see Figure~\ref{GERDA_det}a), for~a total mass of 15.6~kg, were employed together with three non-enriched semi-coaxial detectors. The~semi-coaxial detectors were originally produced by ORTEC for the former \HDM~\cite{hdm2001} and \IGEX~\cite{igex2002} experiments (see Section~\ref{sec:historical_Ge_review}), then refurbished by Canberra and redeployed in the \GERDA\ apparatus. In~July 2012, the natural Ge detectors were replaced by five Broad Energy Germanium (BEGe) diodes~\cite{bege} (see Figure~\ref{GERDA_det}b) with a total mass of 3.6~kg. The~detectors were arranged in four strings, each one housing three (five) semi-coaxial (BEGe) detectors. %and surrounded by a 60$~\mu$m-thin walled copper container called \emph{Mini Shroud} conceived to reduce the collection of $^{42}$K ions onto the surface of the detectors. The~detector volume was enclosed in a 30$~\mu$m-thin copper cylinder of 75~cm of diameter called \emph{Radon Shroud}~\cite{Agostini:2013tek}. 

From September 2013 to December 2015, a major upgrade of the experiment was carried out to improve the sensitivity to the \onbb\ decay  half-life of \gess\ beyond $10^{26}$~yr with a goal exposure of 100~\kgyr~\cite{gerdaupgrade}. The~\GERDA\ Phase II was designed to reach a BI of the order of \pIIbi, thus running in the so-called background-free regime, i.e.,~having less than one background event in the energy region (\qbb\ $\pm~0.5$~FWHM) for the whole exposure. In~the background-free regime,  the sensitivity is expected to scale linearly with the exposure, thus allowing to reach the desired final goal. The~\GERDA\ Phase II detectors array included 40 diodes in total, composed of 7 Phase I semi-coaxial detectors, 30 newly produced BEGes (for a mass of about 20~kg) and 3 non-enriched semi-coaxial detectors.  
%altri upgrade LAr, low mass etc...

The data-taking of \GERDA\ Phase II started in December 2015. From~April 2018 to July 2018, after~having collected an exposure of 58.9~\kgyr~\cite{science}, a~minor upgrade of the experiment was carried out with the installation of five additional inverted coaxial (IC~\cite{cooper2011,gerdaICPC}) detectors (see Figure~\ref{GERDA_det}c), produced in collaboration with Mirion Technologies with a total mass of 9.6~kg. The~data-taking was then resumed and lasted till November 2019 for a final total collected exposure of 103.7~\kgyr~\cite{gerdafinal}. 

\unskip

%GERDA timeline 
%Phase I, major upgrade, Phase II, minor upgrade
%Ge detector employed, characteristics, performances
      
\subsection{Experimental~Setup} \label{GERDA-setup}
The 3500~m.w.e. of rock overburden of the \LNGS\ site, where the \GERDA\ apparatus is installed, provides a reduction of the cosmic muon flux by six orders of magnitude with respect to~surface. 

\begin{figure}[H]
\includegraphics[width=12 cm]{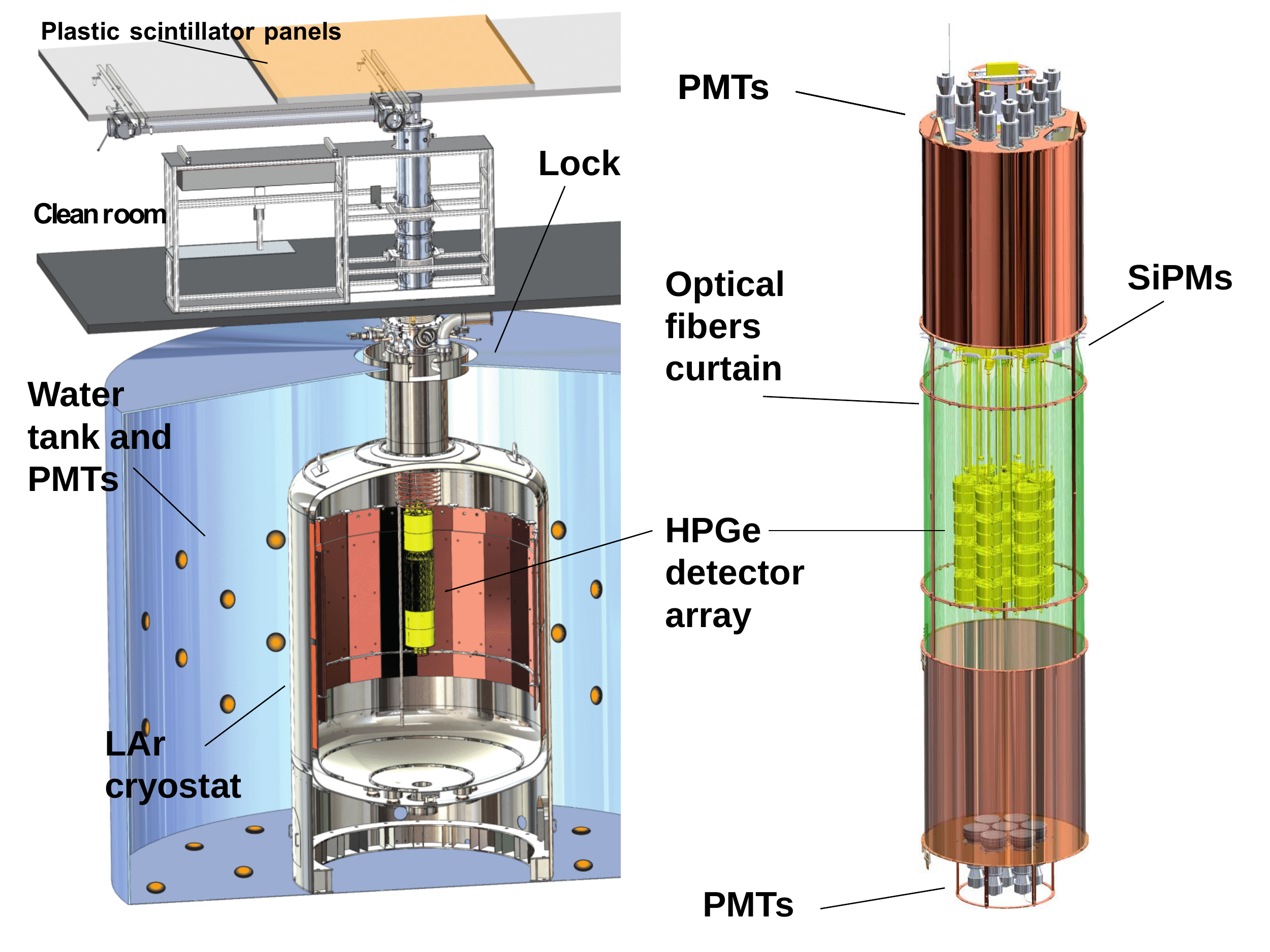}
\caption{Schematic view of the \gerda\ setup. Figure published by Nature, 2017~\cite{nature}. \label{GERDA-layout}}
\end{figure}

The experimental setup consists of a 590~m$^3$ stainless steel tank of 9~m height and 10~m diameter, filled with ultra-pure water~\cite{gerdatech}. The~tank is instrumented with 66 photomultipliers tubes (PMTs), acting as an active Cherenkov veto against the residual cosmic muon flux. The~muon veto system~\cite{Freund:2016fhz} is complemented by scintillator panels installed on the top of the clean room, as shown in Figure~\ref{GERDA-layout}.

%\centering
%\includegraphics[width=12 cm]{figures/GERDA-layout.pdf}
%\caption{\textbf{a}: overview of the \GERDA Phase II setup. 1, water tank; 2, LAr cryostat; 3, clean room; 4, lock; 5, glove box for detector assembling; 6, platsic muon veto system. \textbf{b}: LAr instrumented volume. \textbf{c}: Detector string array. Figure adapted from~\cite{nature}. \label{GERDA-layout}}
%\end{figure} 

The water tank contains a 64~m$^3$ vacuum-insulated stainless steel cryostat (diameter~$=4$~m), filled with LAr. The~internal walls of the cryostat are covered with a 6~cm low-background copper (Cu) layer to shield the $\gamma$ radioactivity originating from the~steel. 

The 40 germanium detectors employed in \GERDA\ Phase II were arranged in seven strings, each one surrounded by a nylon cylindrical vessel conceived to reduce the collection of $^{42}$K ions, originating from the decay of $^{42}$Ar, onto~the surface of the detectors. The~array is lowered in the LAr cryostat from the clean room above the tank through the lock system, as shown in Figure~\ref{GERDA-layout}. 

During the upgrade between Phase I and Phase II, a cylindrical volume around the detector string of about 1~m height and 0.5~m diameter was instrumented with a curtain of wavelength-shifting fibers to detect the LAr scintillation light. The~readout was  performed using 90 silicon photomultipliers (SiPMs) and 16 PMTs, arranged as shown on the right side of Figure~\ref{GERDA-layout}. Both water and LAr  systems act simultaneously as a passive shield from external radioactivity and neutrons and~as active vetoes. This setup allows suppressing the external $\gamma$ background at \qbb\ to less than $10^{-5}$~\ctsper~\cite{Barabanov:2009zz}, while the muon rejection efficiency is $>99.9\%$~\cite{Freund:2016fhz}.

While differing in size and geometry (see Figure~\ref{GERDA_det}), both semi-coaxials and BEGes are p-type semiconductor detectors operated in reverse bias mode. They are fabricated from high purity Ge crystals with an active net impurity concentration of around 10$^{10}$ atoms$/$~cm$^3$~\cite{science}. The~n+ contact is made of diffused lithium with a thickness of about 0.5~mm. The~p+ contact is made of ion-implanted boron with a thickness of the order of 100~$\upmu$m. The~semi-coaxial diodes have a mass of the order of 2~kg and an enrichment fraction ranging from 85.5 to 88.3$\%$~\cite{gerdaupgrade}. The~BEGes detectors have a diameter ranging from 58.3 to 79.3~mm and heights from 22.9 to 35.3~mm~\cite{gerdaupgrade}, with~a mass of the order of O(0.8)~kg. The~enrichment fraction is 87.8$\%$. IC detectors have a mass of the order of 2~kg, a~diameter ranging from 72.6 to 76.6~mm and heights from 80.4 to 85.4~mm. The~enrichment fraction is 88$\%$~\cite{gerdaICPC}.

The different readout electrode layout of semi-coaxial, BEGe and IC is at the basis of their different mass and performance. The~geometry of the p+ contact of a coaxial detector allows the depletion of a larger volume with respect to a BEGe. Conversely, the~latter, thanks to its small p+ electrode, features a lower capacitance (1pF vs. 30 pF), thus resulting in a lower series noise and in a superior energy resolution. This difference is also the key for the better Pulse Shape Discrimination (PSD) performance of BEGe detectors. The~new design of IC detectors allows to increase the detector mass to the level of a coaxial diode, thus lowering the background amount per mass unit while retaining the resolution and the PSD performance of BEGes (see Section~\ref{GERDA-HowTo}).

The readout of germanium detectors is performed using charge sensitive amplifiers located in LAr, 35~cm above the array. The~signal trace, with~a length of 160~$\upmu$s, is sampled at 25~MHz, while a 10~$\upmu$s window around the rising edge is sampled at 100~MHz. Digitized data are stored on a disk for the subsequent~analysis.   

\subsection{Data Analysis Flow and Active Background~Suppression} \label{GERDA-HowTo}
%Calibration
%Efficiencies
%Dead Time
%Blind analysis

%mu-veto, LAr, anti-coincidence, PSD

Since Phase I, \GERDA\ adopted a blind-analysis procedure consisting of removing all the events within $\pm$25~keV of \qbb\ from the analysis flow until all the procedures and the cuts are~finalized.

The stability of detectors, as~well as leakage currents and noise, are monitored through the injection of 0.05~Hz test pulses (TPs). Quality cuts based on the flatness of the baseline, polarity and time structure of the pulse, allow rejecting non-physical events, as~discharge and noise bursts, while keeping 99.9$\%$ of \qbb\ events. 

The event energy reconstruction is performed with a zero-area cusp filter~\cite{zac}. The~energy scale is determined through weekly calibration runs with $^{228}$Th sources. The~energy resolution at \qbb\ in terms of FWHM is (4.9~$\pm$~1.4), (2.6~$\pm$~0.2), (2.9~$\pm$~0.1)~keV for semi-coaxial, BEGe and IC detectors,  respectively~\cite{gerdafinal}. The~energy resolution is stable within 0.1~keV in the whole data-taking~period. 

\begin{figure}[H]
\includegraphics[width=13 cm]{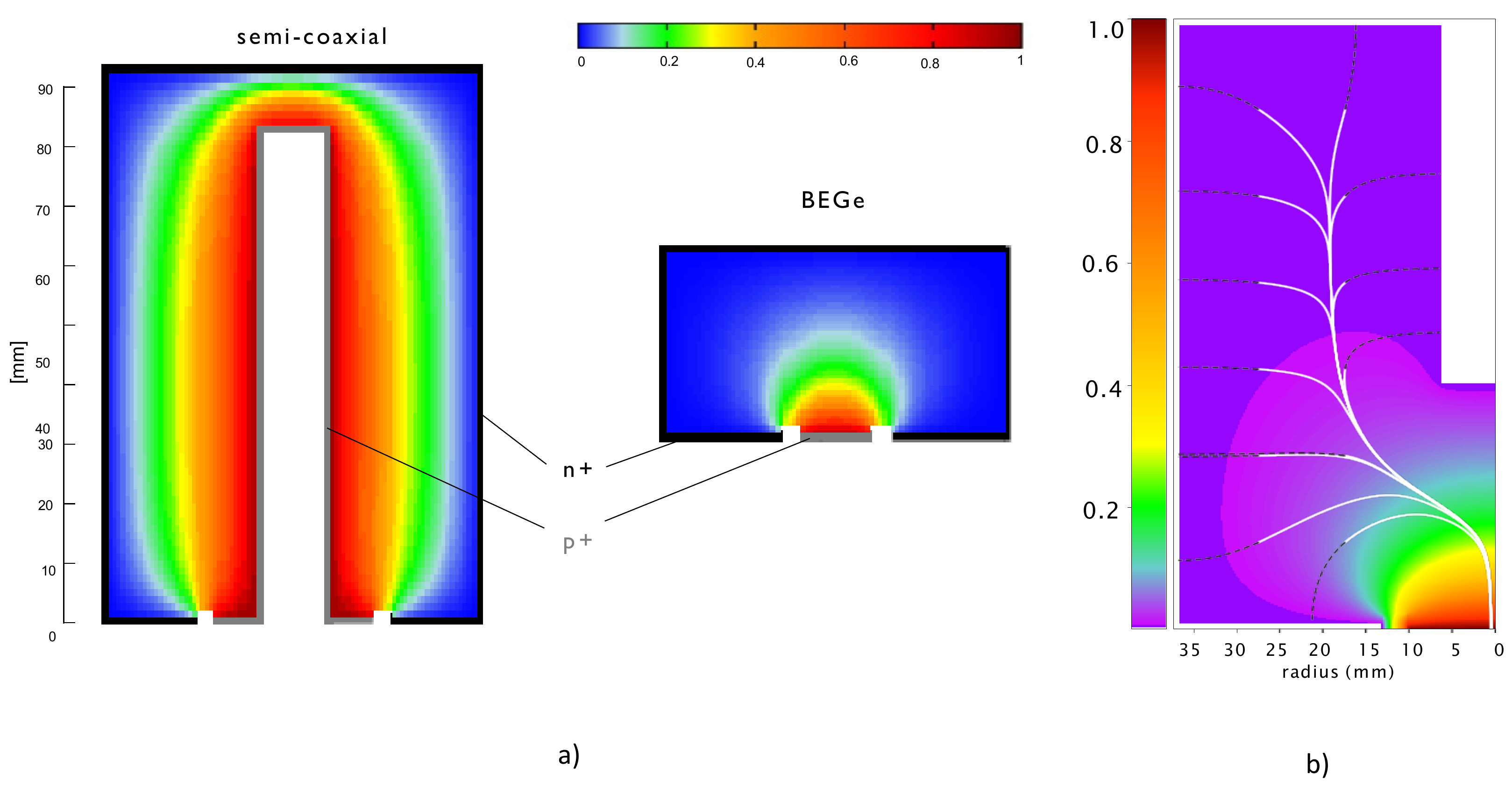}
\caption{(\textbf{a}) Cross-sections of a GERDA coaxial (left) and BEGe (right) detector with an overlay of the corresponding weighting potentials. Figure published by Science, 2019~\cite{science}. (\textbf{b}) Weighting potential for an IC detector. Figure published by EPJC, 2021~\cite{gerdaICPC}.\label{GERDA_det_WeightedPotential}}
\end{figure}

The background suppression is based on different levels. Candidate events within 10~$\upmu$s from a muon veto signal are rejected. The~related dead time is $<$0.01$\%$. In~the same way, events coinciding with an energy deposition in LAr are also vetoed. If~at least one photoelectron within 6~$\upmu$s from a germanium detector trigger is detected by any of the photosensors of the LAr veto system, the~event is classified as background. The~dead time in this case is of the order of 1.8$\%$~\cite{gerdafinal}. Since the range in germanium of the two electrons of a $\beta\beta$ decay event is of the order of 1~mm, the~energy deposition is expected to be highly localized, thus featuring a single-site event (SSE). Conversely, $\gamma$ rays (at~this energy, mainly interacting via Compton scattering) are expected to deposit their energy at multiple sites featuring the so-called multi-site events (MSEs). Anti-coincidence among different Ge detectors is, therefore, exploited to discard background events. Consecutive events occurring within 1~ms are also tagged to veto time-correlated decay from primordial radioisotopes. Finally, the~discrimination between SSEs and MSEs, as~well as from $\alpha$ and $\beta$ surface events, is performed by exploiting the different characteristics of the pulse shape~\cite{Agostini:2013jta}. For~BEGe and IC detectors, the PSD cut is based on the parameter $A/E$, where $A$ is the maximum of the current amplitude and $E$ is the energy. MSEs and n+ surface events show wider current pulses, thus featuring a lower $A/E$ value with respect to SSEs. On~the contrary, surface events on the p+ electrode feature a higher $A/E$~\cite{dusan}. A~mono-parametric cut on both sides of the $A/E$ distribution of SSEs is, therefore, effective at enhancing the signal-to-background ratio. Figure~\ref{GERDA_det_WeightedPotential} shows the weighting potential describing the coupling of the charge with respect to the distance from the respective electrode~\cite{science} for semi-coaxial and BEGe detectors (Figure~\ref{GERDA_det_WeightedPotential}a) and for IC ones (Figure~\ref{GERDA_det_WeightedPotential}b). Due to their different geometries, semi-coaxial diodes show a more complicated pulse time structure requiring the application of an artificial neural network (ANN) to discriminate SSEs from MSEs and an additional cut on signal rise time to reject events on the p+ electrode~\cite{Agostini:2013jta,science}. Both ANN and $A/E$ methods are trained using calibration data. The~\onbb\ signal efficiency is estimated to be (68.8~$\pm$~4.1)$\%$, (89.0~$\pm$~4.1)$\%$, (90.0~$\pm$~1.8)$\%$ for coaxial, BEGe and IC detectors, respectively~\cite{gerdafinal}.   

\unskip 

\subsection{Statistical Analysis and \onbb\ Results} \label{GERDA-results}
The total exposure collected during \GERDA\ PhaseII is 103.7~\kgyr. The~energy range around the \qbb\ considered for the analysis goes from 1930 to 2190~keV with the exclusion of two known background peaks at (2104~$\pm$~5) and (2119~$\pm$~5)~keV. After~the unblinding and the application of the analysis cuts, 13 events are found in the analysis window. The~energy distribution of those events is fitted assuming a flat distribution for background and a Gaussian centered at \qbb\ with a width according to the energy resolution for~a possible \onbb\ signal. Both frequentist and Bayesian analyses are~applied. 

%The energy spectra around \qbb\ for Phase~I, Phase~II coaxial detectors and Phase~II BEGe detectors (after all cuts) are shown in Figure~\ref{fig:roizoom}.

The frequentist analysis is performed using a two-sided test statistics based on the profile likelihood and gives no indication for a signal. The~limit on the half-life of \gess\ is \thalfzero~$> 1.5 \times 10^{26}$~yr (90\% CL). The~combined analysis of the whole Phase I and Phase II data sample, with~a total exposure 127.2~\kgyr, provides the limit \thalfzero~$> 1.8 \times 10^{26}$~yr (90\% CL). The~limit coincides with the sensitivity, defined as the median expectation under the no signal hypothesis~\cite{gerdafinal}.

The background index (BI), as~derived from the fit, reached the unprecedentedly low value of  $BI=5.2^{+1.6}_{-1.3} \times 10^{-4}$~\ctsper in Phase II. The~mean background expected in the signal region (\qbb\ $\pm$ 2$\sigma$) is 0.3 counts, thus reaching the design goal of a background-free ~regime.  

The limit obtained with the Bayesian analysis for Phase I and Phase II data is \thalfzero~$> 1.4 \times 10^{26}$~yr (90\% CL). The~reader is referred to~\cite{nature, science, PRL2018, gerdafinal} for  the different data releases of Phase II and for further details on the statistical~analysis. 

In addition to standard \onbb\ searches, \GERDA\ explored other physics topics such as the search for bosonic superweakly interacting massive particles (super-WIMPs) as keV-scale dark matter candidates~\cite{GERDA:2020emj}, the~search for \onbb\ decay processes accompanied with Majoron emission~\cite{Agostini:2015nwa} and the study of two-neutrino double beta decay of \gess\ to excited states of $^{76}$Se~\cite{Agostini:2015ota}.

\section{The \Majorana\ Experiment} \label{sec:majorana}
The {\sc Majorana} Collaboration searches for neutrinoless double-beta decay of $^{76}$Ge with High Purity Germanium detectors~\cite{Abgrall:2013rze}. The~{\sc Majorana  demonstrator}~\cite{majorana2018} implements an array of 58 HPGe detectors for a total mass of 44.8 kg (14.4 kg of
natural Ge detectors and 29.7 kg enriched to 88.1~$\pm$~0.7\% in $^{76}$Ge), arranged in two different modules, each contained in a low-background shield. The~experiment is located at the
Sanford Underground Research Facility (SURF) in Lead, South Dakota (U.S.A.), at~a depth of \mbox{4300 m.w.e.} Its main goal is to demonstrate a low level of background to justify the construction of a tonne-scale experiment that would probe $m_{\beta\beta}$ at the level of 15~meV with a nearly background-free region around the \qbb. 
%HPGe detectors have very good features for neutrinoless double-beta decay searches, thanks mainly to an excellent energy resolution and to the ability of discriminating among different particles with pulse-shape based algorithms. 
%
%The enrichment  process performed by Majorana achieved an unprecedented yield of 69.8\%~\cite{Abgrall:2017acl} and an energy resolution of 2.53$\pm$0.08 keV at the Q$_{\beta\beta}$.
The enriched detectors are p-type, point contact (PPC) detectors~\cite{34577,Barbeau_2007} with a sub-GeV energy threshold and low capacitance; this allows low-energy physics searches as additional science~channels. 

\begin{figure}[H]
    \includegraphics[width=0.7\textwidth]{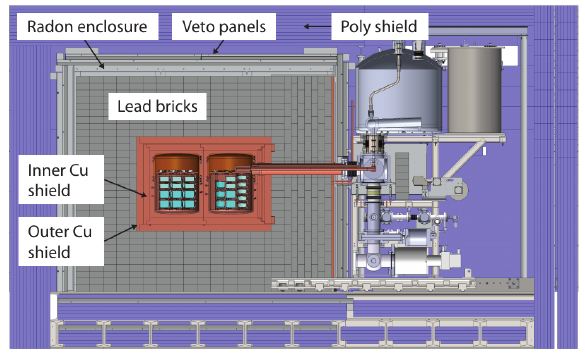}
    \caption{Pictorial view of the Majorana experimental setup. Figure published by J. Phys. Conf. Ser., 2020~\cite{Guinn_2020}.}
    \label{fig:majorana_scheme}
\end{figure}

Two low-background shields contain the two modules of detector arrays. These shields are made of underground electroformed copper (UGEFCu) plus an additional 5~cm of commercial C10100 copper. Outside the copper shields, an additional high-purity lead shielding, 45~cm thick, is enclosed in a region with liquid-nitrogen boil-off gas to reduce the radon contamination. An~active muon veto is located outside the radon exclusion volume and is surrounded by 5~cm of borated polyethylene and 25~cm of polyethylene for neutron moderation.
For each module, energy calibrations are done with $^{228}$Th sources inserted into the shield on a weekly basis~\cite{Abgrall:2017gpr}. A~pictorial view of the \majorana\ experimental setup is shown in Figure~\ref{fig:majorana_scheme}. 
To achieve an ultra-low background, the \Majorana\ {\sc demonstrator} uses a total of 1196~kg of underground electroformed copper, not only for the innermost 5~cm of shielding surrounding the cryostat but also for the cryostats and the detector support structures. In~addition, commercially available low-background materials were carefully screened and selected to be used for cabling, cryostat seals, and~all electrical and thermal insulations. Low-background front-end electronics were developed as well~\cite{Guinn:2015szy}.

\subsection{Data-Taking and Event~Selection}
\label{sec:majorana_HowTo}
Data collected by the \Majorana\ {\sc demonstrator} are divided into seven data sets (DS), from~DS0 to DS6. A~new data set was defined when a significant change in the experimental setup occurred, through the detector setup construction and commissioning. A~letter following the data set number indicated a minor change of the experimental configuration. Data set 0 (DS0) began on 26 July 2015 with the first module of the detector array. The~DS6 data acquired up to 16 April 2018 is labeled DS6a.

Data blindness was implemented through a prescaling scheme in which 31 h
of open background data were followed by  93 h of blind data.
The detector signals are digitized with a 14-bit 100~MHz digitizers~\cite{Abgrall:2020jto}, which were designed for the {\sc GRETINA} experiment~\cite{VETTER2000105}. The~signal waveforms are recorded
in a 20~$\upmu$s acquisition window at the full sampling rate. Each detector had a
high-gain and a low-gain signal amplification and both were digitized independently. The~trigger threshold for each channel was set independently according to the channel trigger
rate, depending on the electronic noise and the
initialization of the on-board trapezoidal filter (see Section~\ref{subsec:energyestimation}). A~reduction in the live time for each detector was estimated at the level of <0.1\% when the initialized value of the triggering filter was negative (because of electronic noise or baseline recovery at the time of initialization due to interactions).

A physical event is represented by recorded waveforms grouped within a 4~$\upmu$s coincidence window. Events in which multiple detectors trigger are rejected. Each waveform is then processed through quality checks to remove non-physical waveforms and signals from periodic pulses. The~acceptance after these quality cuts is estimated to be >99.9\% for all data sets. Events within 1~s from a muon veto trigger are also rejected. Every 36 h, 30~min of data are also rejected for each module due to the filling of liquid nitrogen, which causes microphonic~noise.

\subsection{Energy~Estimation}
\label{subsec:energyestimation}
The energy estimation is done by calibrating the amplitude of the recorded signals once filtered and pole-zero adjusted. 
Corrections that account for ADC non-linearities and charge trapping along the drift path are applied to the acquired waveforms for each respective channel. After~corrections, the~energy uncertainty due to ADC effects is less than 0.1~keV. The energy resolution is also affected by drift-path-dependent charge trapping in the crystal bulk. This effect is taken into account with an additional term for the standard pole-zero adjustment:
\begin{equation}
    \frac{1}{\tau} = \frac{1}{\tau_{PZ}} + \frac{1}{\tau_{CT}}
\end{equation}
where $\tau_{PZ}$ is the pole-zero time constant due to the preamplifier, and $\tau_{CT}$ is the correction that reproduces exponential trapping of charges on the drift path. For~each detector, the total pole-zero correction $\tau$ is optimized by minimizing the full-width at half maximum (FWHM) of the 2615~keV $^{208}$Tl peak in calibration data: this correction improves the energy resolution of 1.4~keV on average. 
A fast trapezoidal filter (with a rise time of 1.0~$\upmu$s and a flat-top time of 1.5~$\upmu$s) is applied to the waveform to estimate the start time $t_0$ and the threshold crossing time. Then the pole-zero correction and a slower trapezoidal filter (with a rise time of 4.0~$\upmu$s and a flat top time of 2.5~$\upmu$s) are applied to the original waveform. 
The energy of the event is estimated as the value of the waveform at a time of 0.5~$\upmu$s from the end of the flat top, relative to t$_0$, after~having applied the above corrections and~filtering.  

Periodic energy calibrations are used to provide an initial linear energy scale calibration for each channel, taking into account small variations over time of the electronic noise or energy scale~\cite{Abgrall:2017gpr}. Then, the~combination of the spectra obtained provides a more precise energy calibration through the simultaneous fit of the full energy calibration peaks at 239~keV ($^{212}$Pb), 241~keV ($^{224}$Ra), 277~keV ($^{208}$Tl), 300~keV ($^{212}$Pb), 583~keV ($^{208}$Tl), 727~keV ($^{212}$Bi), 861~keV ($^{208}$Tl), and~2615~keV ($^{208}$Tl), respectively. 

\begin{figure}[H]
     \includegraphics[width=.7\textwidth]{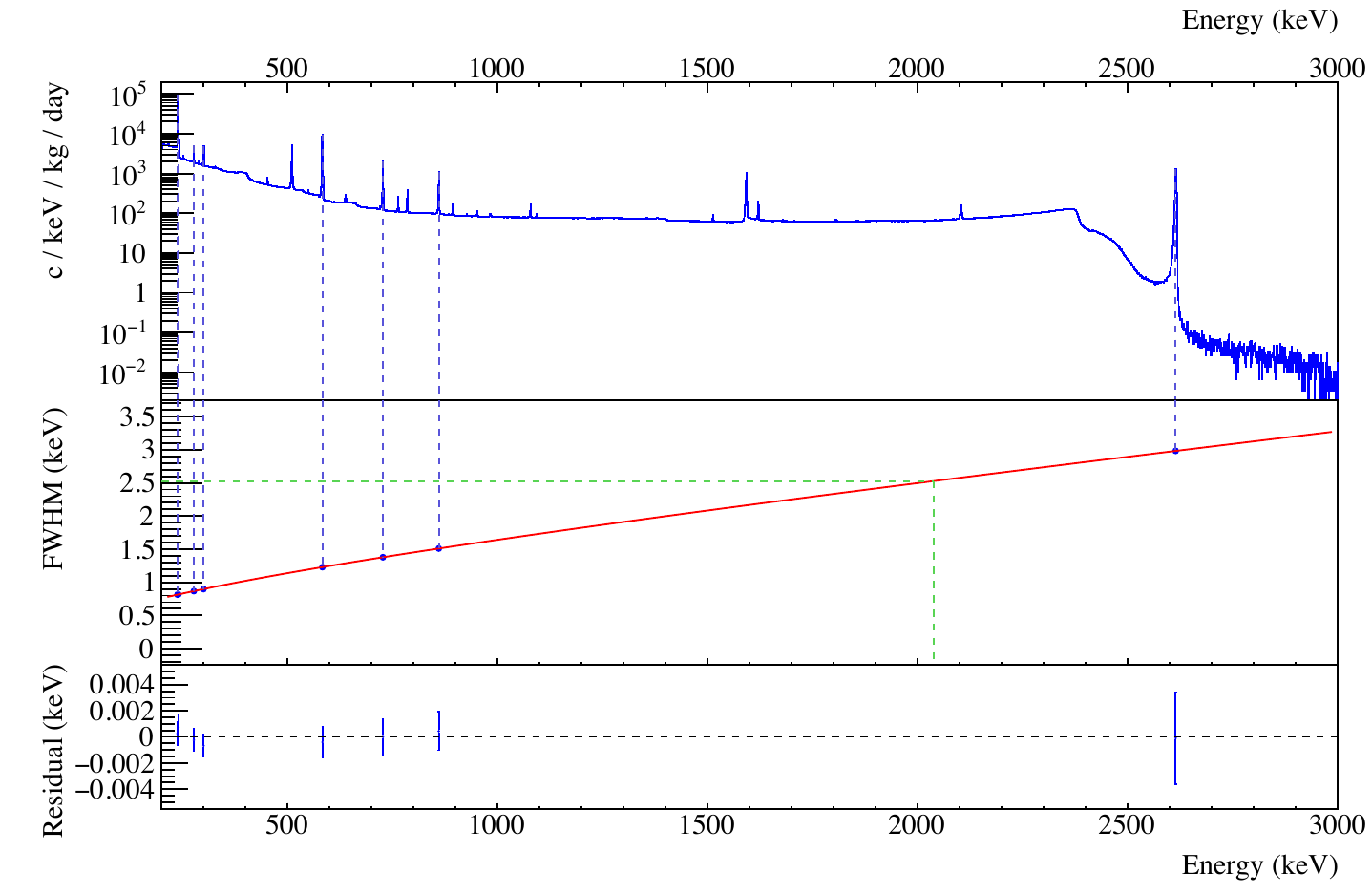}
    \caption{\textbf{Top}: Calibration spectrum from combined data of DS0 to DS6. \textbf{Center}: exposure-weighted resolution for all the calibration peaks and the FWHM fit function (red line). The~green line indicates the exposure-weighted resolution value at 2039~keV. \textbf{Bottom}: residuals of the estimated FWHM values from the fit. Figure published by Phys. Rev. C, 2019~\cite{majorana2019}}
    %MDPI: Please moved Figure 9 after it first citation and make sure the order of references is correct.
    \label{fig:majorana_calibration}
\end{figure}

Each peak is fitted with a response function $R(E)$ given by the sum of a Gaussian function and an exponentially modified Gaussian Tail:
\begin{equation}
    R(E)  = \frac{1-f}{\sqrt{2 \pi \sigma^2}} e^{-\frac{(E-\mu)^2}{2 \sigma^2}} + \frac{f}{2 \gamma} e^{\left(\frac{\sigma^2}{2 \gamma^2} + \frac{E - \mu}{\gamma}\right)} \mbox{erfc} \left(\frac{\sigma}{\sqrt{2}\gamma} + \frac{E - 
    \mu}{\sqrt{2} \sigma}\right)
\end{equation}
where $\sigma$ represents the smearing due to electronic noise and partial charge collection, $\gamma$ is the decay constant of the low-energy tail, and~$f$ is the fraction of the peak shape contained in the low-energy tail.
The background in the surrounding region of the peak is modeled by the sum of an error function and a continuum component given by a quadratic polynomial.
The combined calibration spectrum from DS0 to DS6 is shown in the top panel of \mbox{Figure~\ref{fig:majorana_calibration}.} After~having determined the FWHM at each energy peak, the~FWHM is fitted with the following function of the energy $E$:
\begin{equation}
    FWHM(E) = \sqrt{\Gamma_n^2 + \Gamma_F^2 E + \Gamma_q^2 E^2}
\end{equation}
where $\Gamma_n$, $\Gamma_F$ and $\Gamma_q$ are the terms due
to the electronic noise, the~Fano factor~\cite{PhysRev.72.26}, and~the partial charge collection, respectively. The~central panel of Figure~\ref{fig:majorana_calibration} shows 
the exposure-weighted resolution for each gamma peak and a fit to the exposure-weighted values. The~exposure-weighted average resolution (FWHM) at Q$_{\beta \beta}$ is (2.53 $\pm$ 0.08) keV.

\subsection{Background~Suppression}
\label{sec:majorana_bkg_suppression}
\begin{figure}[H]
    \includegraphics[width=0.35\textwidth]{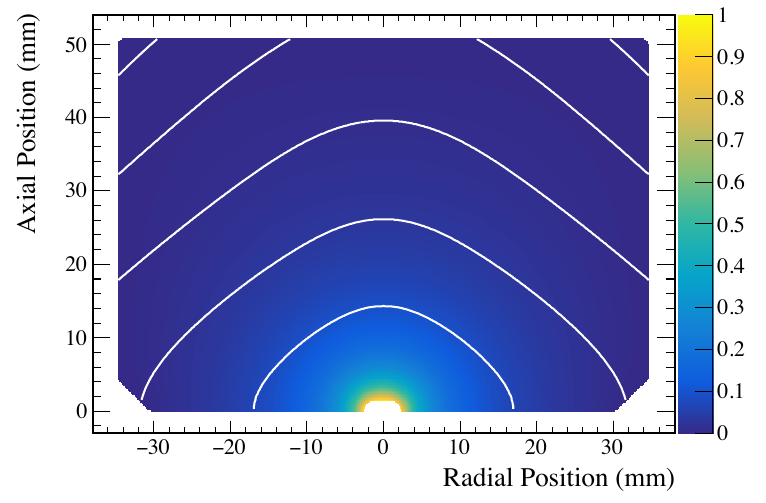}
    \includegraphics[width=0.35\textwidth]{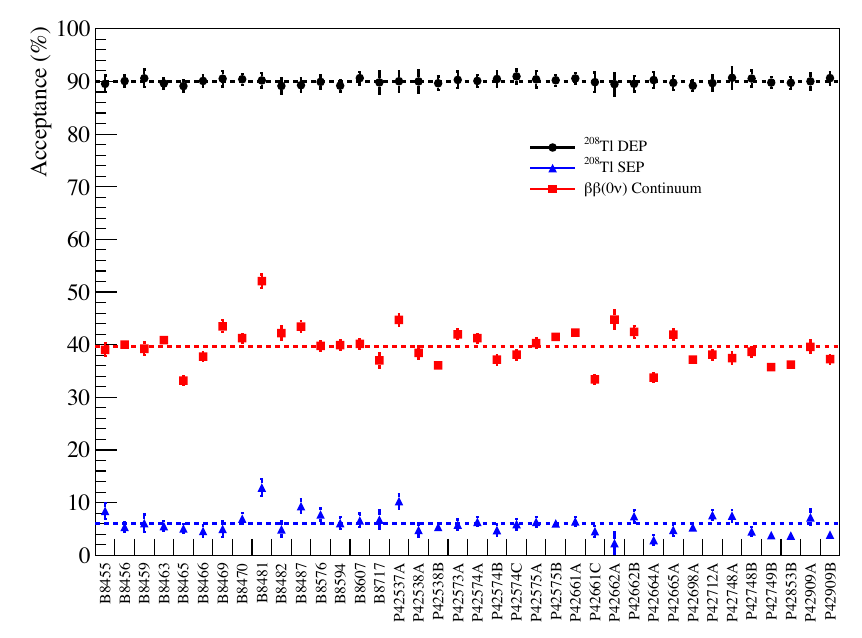}
    \caption{\textbf{Left}: weighting potential of the PPC detectors. White lines indicate equal drift times. \textbf{Right}: acceptance for each detector in calibration data for DS6a. Horizontal lines represent the mean values for all calibrated detectors. Figure published by Phys. Rev. C, 2019~\cite{majorana2019}}
    \label{fig:potential_ppc_and_AoverE}
\end{figure}

The weighting potential of PPC Ge detectors is relatively smaller in the bulk of the crystal and
mostly located in the vicinity of the point contact (see Figure~\ref{fig:potential_ppc_and_AoverE}, on the left). As~also explained  in Section~\ref{GERDA-HowTo}, electrons interacting inside the bulk of the detectors are identified as single-site events (SSEs), as their range is limited to less than 1 mm at the energies of interest. Gamma rays, instead, interact inside the detectors mostly as multiple-site events (MSEs). This translates to a difference in the pulse shape, which allows the discrimination of the gamma-ray background. In~particular, at~about the same energy ($E$), multi-site events have a maximum current amplitude ($A$) quite smaller than single-site events; calibration data are therefore used to fit the mean value of $A$ as a function of $E$ for each data set and detector. The~parameter AvsE is defined as:
\begin{equation}
    \mbox{AvsE} = \frac{1}{j} (p_0 + p_1E + p_2E^2 - \lambda A)
\end{equation}
where $p_0$, $p_1$ and $p_2$ are the coefficients obtained from the fit of $A$ as a function of $E$ and $\lambda$ is the calibration constant usedto convert the ADC channels to energy expressed in~keV.

The value of AvsE is then used to establish a cut above $\sim$1 that would have a 90\% efficiency in accepting single-site events from the 1593~keV double-escape peak (DEP) from the $^{208}$Tl gamma ray line at 2615~keV. The~survival percentage for events in the DEP, the~single-escape peak (SEP) and the Compton continuum in a 100~keV region around the Q$_{\beta\beta}$ are shown on the right part of Figure~\ref{fig:potential_ppc_and_AoverE} for events from DS6a and for each detector: 40\% of Compton events are accepted while about 6\% of multi-site events from the SEP are retained. The~main systematic uncertainties in the AvsE estimation are due to the difference between the position distribution of simulated events and the interactions from the calibration sources, the~time-variation in the SSE acceptance and the energy dependence of the cut acceptance~\cite{PhysRevC.99.065501}.

External $\alpha$ particles with MeV energies have a range of tens of $\upmu$m in Ge detectors. The~PPC Ge detectors of \Majorana\ have lithiated dead layers (1.1 mm thick) over the surface and a passivated surface on the face with the point contact. As the dead layers are much thicker  than their range, $\alpha$ particles impinging the lithiated surfaces cannot  penetrate inside the active volume. On~the contrary, they can penetrate the passivated surface and deposit their energy in the active region of the detectors. The~holes near the passivated surface are trapped and later released in a time that is much longer than the rise time of events occurring in the bulk. This degrades the measured energy for $\alpha$ particles, contributing to the background near the Q$_{\beta\beta}$ value but the difference in rise time with respect to events occurring in the bulk allows for their discrimination. A~cut is defined, based on the slope $\Delta$ between the average value of the two 1~$\upmu$s-wide regions, which start either 2~$\upmu$s after the time the waveform reaches 97\% of its maximum (t$_{97}$) or 1~$\upmu$s before the end of the waveform (t$_{\mbox{max}}$). The~parameter DCR (Delayed Charge Recovery) is defined as the following:
\begin{equation}
    DCR = \frac{\int_{t_{max} - 1 \mu s}^{t_{max}}{V(t)dt} - \int^{t_{97} + 3\mu s}_{t_{97} + 2 \mu s}{V(t)dt}}{t_{\mbox{max}} - t_{97}}- \Delta
\end{equation}
and ensures 99.9\% acceptance of the Compton events from calibration data near the \qbb\ value. The~acceptance of the cut is estimated  for each data~set.%Please confirm intended meaning is retained.

\subsection{\onbb\ Search~Results}
\label{sec:majorana_results}

The characteristics and the efficiencies values are  evaluated for each data set. In~addition to the efficiencies for background rejection $\epsilon_{AvsE}$ and $\epsilon_{DCR}$, the~efficiency $\epsilon_{cont}$ of detecting the full energy of \onbb\ events is combined together with the previous ones. The~total efficiency weighted with the exposure of all data sets is 0.810$^{+0.031}_{-0.032}$.  
The \onbb\ ROI
is optimized for each data set using the peak shape parameters and background index. $\epsilon_{res}$ represents the fraction of \onbb\ events falling inside the optimal ROI for each data set. The~exposure weighted value for $\epsilon_{res}$ over all data sets is 0.900 $\pm$ 0.007. 

\begin{figure}[H]
\widefigure
   \includegraphics[width=0.35\textwidth]{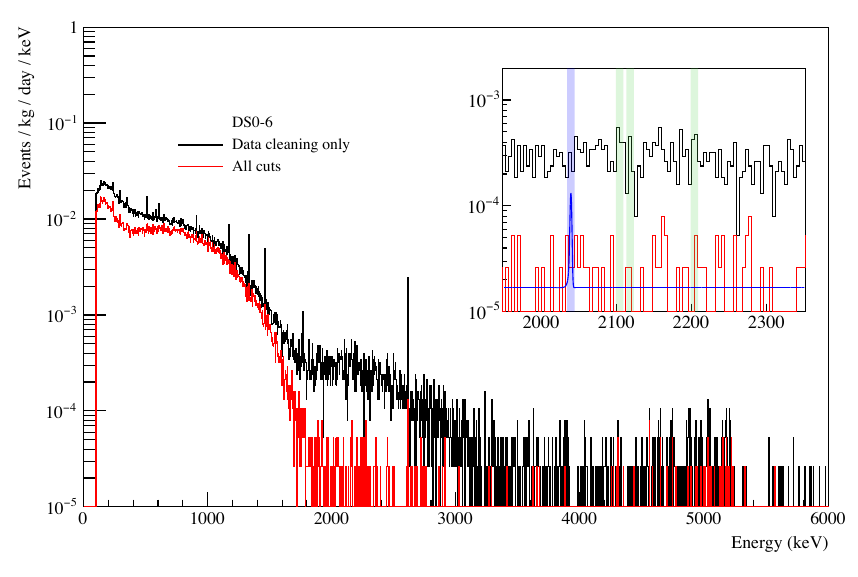}
    \includegraphics[width=0.35\textwidth]{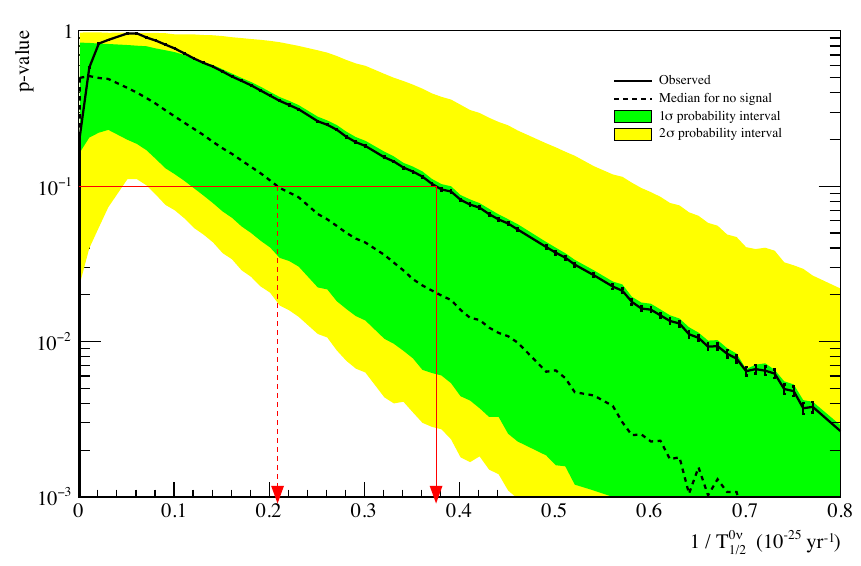}
    \caption{\textbf{Left}: energy spectrum above 100 keV with only data cleaning and muon veto cuts (black) and after all cuts (red). The~blue curve shows the fitted background from the unshaded regions in the inset and the 90\% C.L. upper limit on the counts from a \onbb\ signal.~\textbf {Right}: the $p$-value as a function of the \onbb\ half-life estimated from the unbinned frequentist profile-likelihood method (solid black). The~median sensitivity is shown as a dashed black line, and~the shaded bands represent  the 1 and 2$\sigma$ intervals. Figure published by Phys. Rev. C, 2019~\cite{majorana2019}}
    %MDPI: Please change hyphen to minus sign in image unit.
    \label{fig:majorana_spectrum_results}
\end{figure}

A total of 26 kg$\cdot$yr exposure was collected from DS0 to DS6 of which 11.85 kg$\cdot$yr were blinded across the energy spectrum. The~energy spectrum above 100 keV relative to the full enriched detector exposure is shown in Figure~\ref{fig:majorana_spectrum_results} (left), with~only data cleaning and muon veto cuts (in black) and with all cuts applied (in red). The~inlet shows the background events between 1950 and 2350 keV, where a flat component is expected (from background simulations with the {\sc MaGe} software~\cite{5876017} based on {\sc Geant4}~\cite{Agostinelli:2002hh}) with the exclusion of $\pm$5~KeV of the 2103 keV (single-escape peak from $^{208}$Tl), 2118 keV and \mbox{2204 keV} (gammas from $^{214}$Bi) peaks. The~predicted background in this region is (6.1 $\pm$ 0.8) $\times$ 10$^{-3}$ counts/(keV kg yr) in the  exposure weighted optimal ROI of 4.13~keV and is consistent with 
$^{208}$Tl contamination in components larger than assay values, whose origin is under~investigation. 
\end{paracol}
\begin{paracol}{2}
\switchcolumn

The limit on the half-life of the \onbb\ decay of $^{76}$Ge (considered as a Poisson process) is estimated with a Feldman--Cousins approach (see Reference~\cite{PhysRevD.57.3873}) with 0.65 expected background events and 1 event observed in the ROI at 2040 keV. The~value of the 90\% C.L. lower limit is 
2.5 $\times$ 10$^{25}$ yr. Based on an unbinned, extended profile likelihood method based on RooStats (see References~\cite{Verkerke:2003ir,Schott:2012zb,nature}),  the~median sensitivity at 90\% C.L. is 4.8 $\times$ 10$^{25}$ yr, as shown in Figure~\ref{fig:majorana_spectrum_results}. The~
$p$-value distribution of the half-life is
T$^{0\nu}_{1/2}$ $>$ 2.7 $\times$ 10$^{25}$ yr at 90\% C.L. In~this analysis, the half-life is a common parameter for all data sets, while the
peak shape parameters and signal efficiencies are constrained as a Gaussian nuisance term for each data set. The~corresponding upper limit on the number of signal events at 90\% C.L. is 3.8. A~Bayesian statistical analysis was also performed using Markov chain Monte Carlo simulations
in RooStats with the same likelihood function.
With a flat prior on $1/T^{1/2}_{0\nu}$,
the Bayesian limit on the half-life is 2.5 $\times$ 10$^{25}$ yr for a 90\% credible~interval. 

Based on a range of nuclear matrix elements for $^{76}$Ge (\cite{MENENDEZ2009139,Horoi:2015tkc,Hyvarinen:2015bda,Vaquero:2014dna,Yao:2014uta}), phase-space factors of 2.36 $\times$ 10$^{-15}$/yr (\cite{Kotila:2012zza}) or 2.37 $\times$ 10$^{-15}$/yr (\cite{Mirea:2015nsl}), and~a value of g$_{A}$ = 1.27, the limit on the half-life is converted in a range of upper limits for the effective  neutrino mass given by 
%$\left< 
$m_{\beta\beta}$
%\right>$ 
$<$ (200--%MDPI: We change hyphen to en dash, please confirm.
433) meV~\cite{majorana2019}.

\section{The LEGEND~Project} \label{sec:legend}
Building upon the success of the \GERDA\ and \Majorana\ experiments, the~\legend\ (Large Enriched Germanium Detector for Neutrinoless $\beta\beta$ Decay) Collaboration~\cite{legend} aims at building a \gess-based neutrinoless double beta decay experiment with a sensitivity of the half-life beyond $10^{28}$ years, to~fully span the inverted neutrino mass ordering~region. 

\Majorana\ and \GERDA\ have already proven to have the best energy resolution (see Section~\ref{subsec:energyestimation}), the~lowest background index and the best sensitivity in the field (see Section~\ref{GERDA-results}). 
The \legend\ experiment will benefit from the knowledge and from the technological achievements of the two Collaborations, as~well as from the contributions of new joining~groups.

\legend\ will inherit the shielding concept implemented in \GERDA\ with the use of a water Cherenkov veto and the depletion of bare germanium detectors in an instrumented LAr volume simultaneously acting as a coolant medium and active veto. The~low-noise readout electronic developed by the \Majorana\ Collaboration has proven to be successful in improving the PSD performances as well as in lowering the threshold thus resulting in an unprecedented low energy resolution. Moreover, the~careful selection and radiopurity control of the employed materials allowed \Majorana\ to reach a similar background level with respect to \gerda. 

The natural steps toward reaching higher sensitivities consist of increasing the detector mass while further reducing the background, aiming at performing a background-free measurement at larger~exposures.  

The \legend\ project will proceed in two steps: in the first phase, 200 kg of enriched germanium detectors will be deployed in the existing \gerda\ facility at \LNGS. By~reducing the background index of about a factor of three with respect to the \gerda\ final level (i.e., from \mbox{1.5 cts$/($FWHM$\cdot$t$\cdot$yr)} to \mbox{0.5 cts$/($FWHM$\cdot$t$\cdot$yr)}) and with an exposure of 1 t$\cdot$yr, \legend-200  will be able to reach a sensitivity of about $10^{27}$ yr at $90\%$ C.L. The~data-taking is expected to start by the end of~2021. 

In the second phase, the~enriched germanium mass will be increased up to 1000~kg. By~lowering the background to \mbox{0.025 cts$/($FWHM$\cdot$t$\cdot$yr)} and with an exposure of 10 t$\cdot$yr, \legend-1000 will be able to  reach a $3\sigma$ half-life discovery sensitivity of $1.3\times10^{28}$~yr (see Figure~\ref{discovery}). The~location of the \legend-1000 phase will be selected in order to keep the cosmogenic activation background as low as possible and within the required background~level.

In the next sections, we will present the project and we will discuss the main issues addressed by the \legend\ Collaboration.

\begin{figure}[H]
     \includegraphics[scale=0.55]{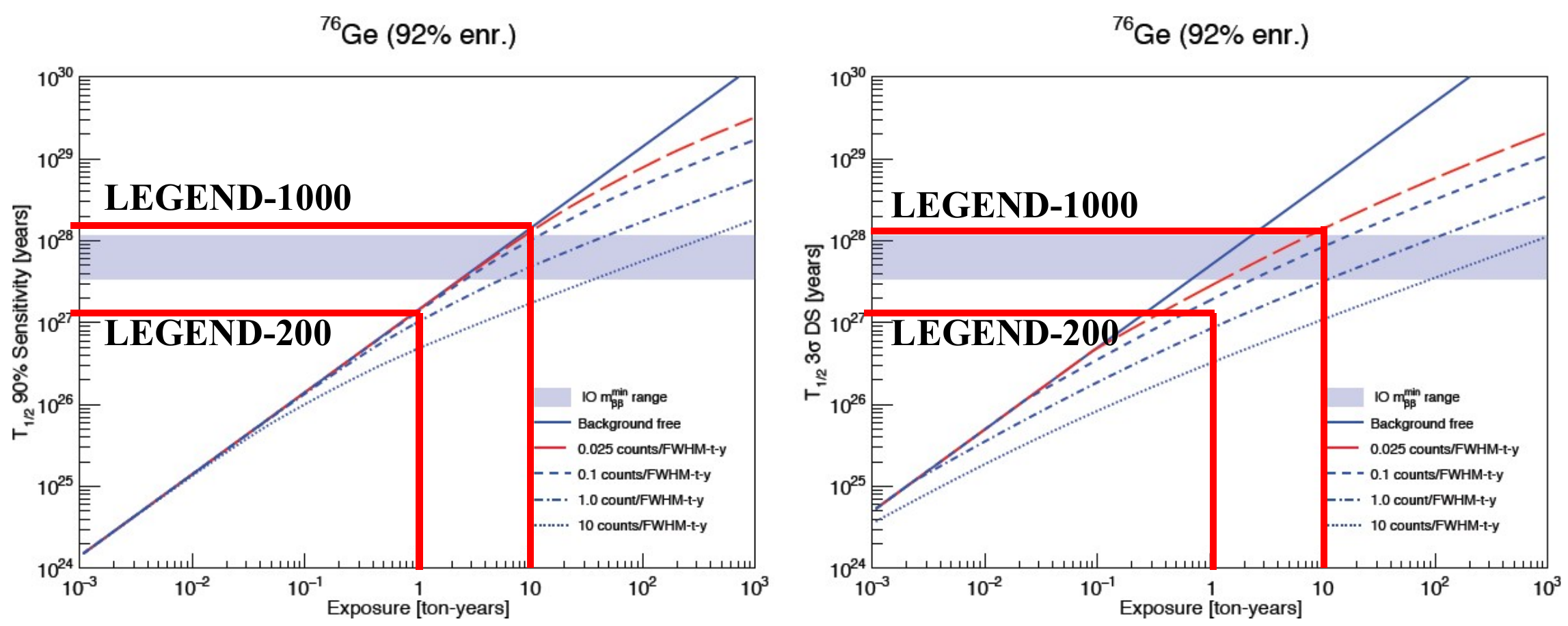}
    \caption{Sensitivity as a function of exposure and background for an isotopic enrichment fraction in \gess\ of $92\%$ and for (\textbf{left}) a 90$\%$~C.L. limit setting and (\textbf{right}) a 3$\sigma$ signal discovery. 
    %The blue band represents the bottom of the inverted neutrino mass ordering region, computed using $m_{\beta\beta}=17$ eV, no axial coupling quenching and nuclear matrix element ranging from 3.5 to 5.5. 
    %Discovery sensitivities are reported for \Majorana\ and \GERDA\ combined together and for the two phases {\mbox{\textsc{Legend}}}-200 (using a background index of 0.5 counts$/$(FWHM$\cdot$t$\cdot$yr) at 1 t$\cdot$yr exposure) and {\mbox{\textsc{Legend}}}-1000 (using  a background index of 0.025 counts$/($FWHM$\cdot$t$\cdot$yr) at 10 t$\cdot$yr exposure). b) Sensitivity. In case of exactly zero background, the sensitivity scales linearly with exposure. 
    Plots from~\cite{Legend1000_pCDR}. }
    \label{discovery}
\end{figure}
\unskip

%\subsection{Background reduction plans}
%\subsection{Inverted Coaxial Point Contact detectors}
\subsection{\legend-200 Germanium Detectors and Experimental~Setup}
The existing \GERDA\ cryostat is large enough to accommodate up to 200 kg of detectors divided into 19 strings for a total diameter of 500--550 mm (see Figure~\ref{legend_setup}a) \cite{legend}. The~approach of reusing the \gerda\ experimental apparatus will assure a timely start of data-taking with a world-leading sensitive~experiment.  

\begin{figure}[H]
    \includegraphics[scale=0.5]{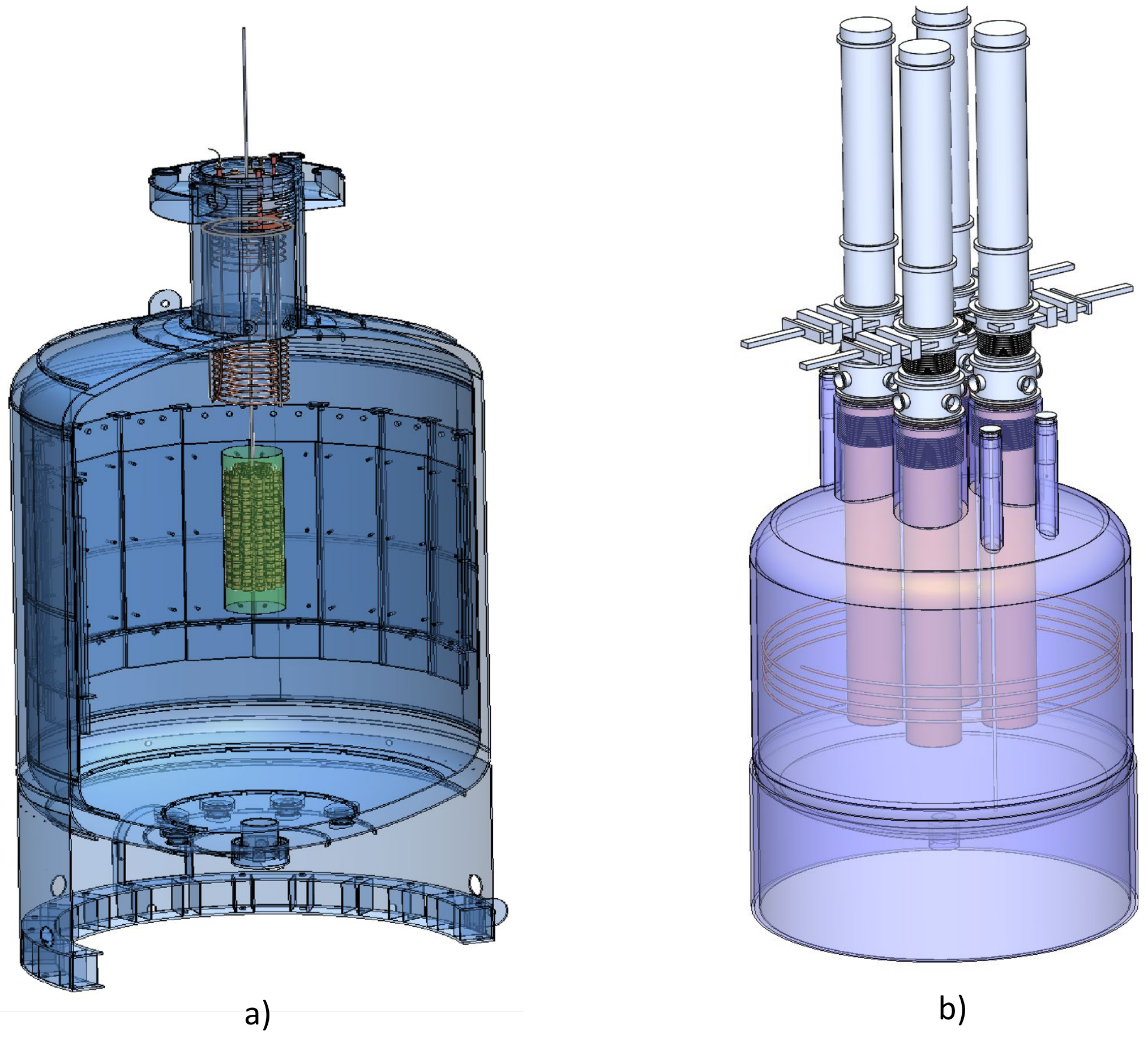}
    \caption{Experimental setup of the \legend-200 (\textbf{a}) and \legend-1000 (\textbf{b}) phase, respectively.}
    \label{legend_setup}
\end{figure}

\legend-200 will deploy 20~kg of BEGe and 9.4~kg of ICPC detectors (see \mbox{Figure~\ref{GERDA_det}b,c}) from \GERDA\, and~28~kg of PPC detectors from \Majorana. In~addition, nearly 140~kg of newly produced ICPC detectors will be added. As~discussed in Sections~\ref{GERDA-setup} and~\ref{GERDA-HowTo}, ICPCs feature a new geometry with respect to the previously used germanium detectors. As~shown in Figure~\ref{GERDA_det}c, a~small p$^+$ electrode is placed on the opposite face with respect to the bore hole and the n$^+$ outer contact covers all the remaining surfaces (cylindrical part and bore hole). Like semi-coaxial detectors, ICPCs can be manufactured with a larger mass (\mbox{$\sim$2 kg}) with respect to BEGes ($\sim$0.8 kg) and PPCs ($\sim$0.85 kg). This allows increasing the active mass while reducing the amount of nearby materials contributing to the background, such as cables, electronics and holders. Moreover, the~reduced surface-to-volume ratio with respect to smaller detectors, makes ICPCs less susceptible to surface effects. Furthermore, because~of the long drift time inside the crystal and the small p$^+$ electrode, the~energy resolution and the pulse shape characteristics are very similar to BEGe and PPC~\cite{domula2018}, making this new type of HPGe suitable for \onbb\ decay~experiments. 

%Characterization of inverted coaxial \gess~detectors in \gerda.
As explained in Section~\ref{GERDA-HowTo}, during~the 2018 upgrade, five \gess\ enriched ICPC detectors were deployed in the \gerda\ cryostat and operated in LAr for the first time until the end of the data-taking in 2019. After~being tested and characterized, they showed energy resolution and background rejection capability comparable with BEGes while having a larger mass by a factor of three~\cite{gerdaICPC,gerdafinal}.

\subsection{Readout~Electronics}
\legend-200 will use a resistive-feedback charge-sensitive amplifier (CSA) operated at cryogenic temperature in LAr. CSA is divided into two stages in order to meet the background requirements. The~first stage is the low mass front end (LMFE~\cite{Abgrall:2015hna}), a~custom-made low-background circuit based on the front-end used in \Majorana\ where a junction gate field-effect transistor (JFET) is installed. LMFE is made of ultra-clean materials (<$1\,\upmu$Bq/kg) due to its close vicinity to the detectors. The~second stage is a differential amplifier $\sim$30--150~cm away from the first stage; in this case, a slightly higher activity (\mbox{50 $\upmu$Bq}) can be tolerated. The~amplifier is based on the one already used in \gerda\ ~\cite{Riboldi:2012ola} and the two stages are connected using four custom-made low-mass coaxial-cables.
The CSA is designed to have electronic noise $<$1~keV~FWHM, an~energy resolution of $\leq$2.5~keV at \qbb, a~fast rise time of $\leq$100 ns to allow powerful pulse shape discrimination analysis and high linearity up to 10~MeV for alpha particles rejection~\cite{Willers:2019vlx}. 

For \legend-1000, in~order to have the same performance with lower background contribution, an~ASIC preamplifier to be placed near the detectors is being~studied.

\subsection{Background Mitigation~Techniques}
The electroformed copper (UGEFCu) employed in the \Majorana\ experiment (see Section~\ref{sec:majorana}), shows a uranium and thorium decay chain activity of  <$0.1\,\upmu$Bq/kg~\cite{Abgrall:2016cct}. This value is at least one order of magnitude lower than that of the most commercial oxygen-free, high-conductivity (OFHC) copper used in the external layers of the \Majorana\ shielding. 
%or of the one employed in \GERDA's detector holders. 
UGEFCu is produced at Pacific Northwest National Laboratory and SURF, where it is also stored underground to minimize the possibility of cosmogenic activation producing $^{60}$Co, whose decay yields photons above the Ge \qbb\ value~\cite{Myslik:2018vts}. The~employment of electroformed copper is going to improve the \GERDA\ radiopurity for \legend-200.

%\subsubsection{LAr veto system}
As introduced above, in~both \legend~phases, HPGe detectors will be operated in a LAr-filled cryostat. LAr acts as a coolant for germanium detectors and a passive shield against external radiation. Moreover, an~active veto system in \GERDA\ was used to discard background events (such as neutrons and photons) depositing energy in LAr and making it scintillating in coincidence with a signal in the Ge diodes. As~shown in Figure~\ref{legend_LAr}, \legend-200 will use a similar design, although with a different geometry, to deal with the increased number of detectors in the array  while maximizing the light collection efficiency. Further enhancement of the light yield and attenuation length can be reached by improving LAr purity and by applying Xe doping~\cite{Myslik:2018vts}.

\begin{figure}[H]
    \includegraphics[scale=0.53]{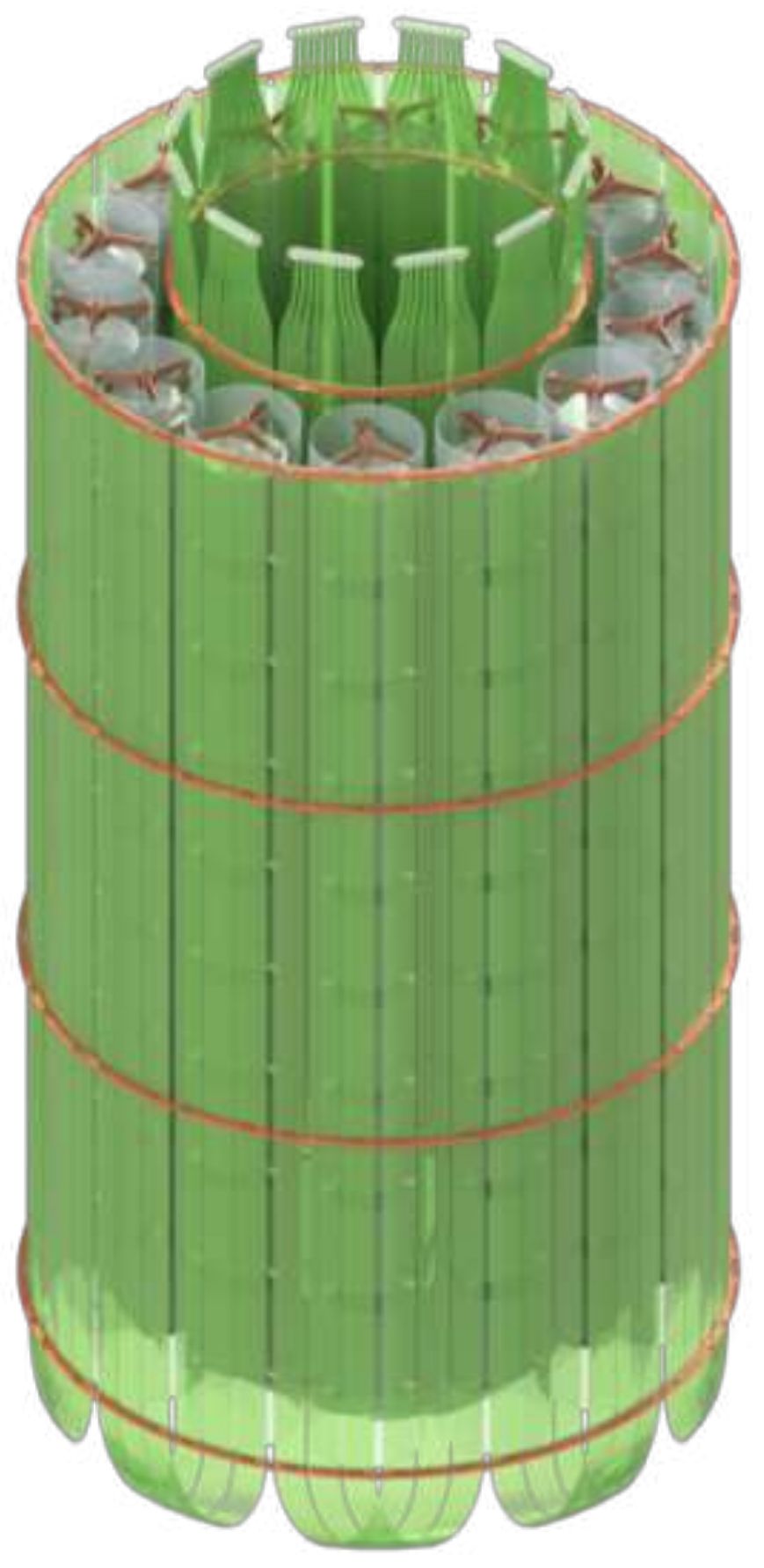}
    \caption{Instrumentation of the LAr volume around the detectors strings in \legend-200.}
    \label{legend_LAr}
\end{figure}

%The luminescence light of LAr is peaked at 128 nm, in the vacuum ultraviolet region (VUV) and therefore a wavelength shifting material is needed (scintillating fibers or tetraphenyl butadiene TBD) to convert this light into near-visible one (about 400 nm) and detect it with standard photodetectors. LAr light yield and attenuation can be improved by increasing argon purity or by adding small amount of xenon, which is currently under investigation.
%Since LAr is only partially transparent to 128 nm photons and they cannot be detected directly but need to be shifted to higher wavelengths, 
%studies are currently ongoing to address the feasibility of adding small amount of Xe impurities to LAr in {\mbox{\textsc{Legend}}}.

%In fact, when energy deposition occurs and Ar excimers form, they can transfer their excitation energy to nearby Xe atoms through collisions and excite them, leading to the formation of Xe excimers that eventually will decay emitting 175~nm photons. LAr is transparent at this wavelength and this shift occurs already at small concentrations of Xe; also, Xe-doped LAr exhibits better pulse shape discrimination performance with increasing Xe concentration with respect to pure LAr.

The use of detector holders made from polyethylene naphthalate (PEN) will further improve the background rejection capability. PEN is a thermoplastic material with scintillating properties and an emission spectrum that peaks in the blue region at 425 nm. Its scintillation yield is 2.5 times lower than standard plastic scintillators, but it has other appealing features such as its very favorable mechanical properties at cryogenic temperature (77 K) and its radiopurity. Moreover, PEN shifts the LAr 128~nm photons to a higher wavelength, making it detectable  with standard photodetectors~\cite{Efremenko:2019xbs}.
%Therefore PEN could be used to build low-background scintillating  structural components of detectors holders, 
By replacing the Si-made, nontransparent detector holders previously used in \gerda with PEN supports,  it would be possible to improve the light collection in the vicinity of the detectors while at the same time exploiting PEN self-vetoing capabilities~\cite{Abt:2020pwk}. The~employment of Pulse Shape Discrimination techniques with PEN is also feasible~\cite{Efremenko:2019xbs}. In~this way, it could be possible to more efficiently reduce the background due to surface events and detect LAr scintillating light close to the detectors. PEN holders specifically designed and optimized to reduce the total mass in contact with the detectors are being produced and will be employed in \legend-200~\cite{Abt:2020pwk}.

An important source of background in \legend-200 is the $^{42}$Ar, a~cosmogenically produced Ar isotope that $\beta$-decays to $^{42}$K. The~distribution of $^{42}$K in LAr is quite likely to be inhomogeneous due to its drift in the electric field generated by high-voltage cables and detectors and also convective motion. Because~of its very high $\beta^-$ $Q$-value of $3.5$ MeV, higher than \qbb, $^{42}$K contributes to the backgrounds in \onbb\ searches. In~\legend-200, the drift of $^{42}$K towards germanium detectors will be reduced by nylon shrouds around each detector's columns (following the \gerda\ approach). \legend-1000, because~of more stringent background requirements, will make use of underground Ar, which is free in $^{42}$Ar, thus almost completely removing the $^{42}$K~background.

In addition to the aforementioned techniques, \legend\ will follow the \gerda\ and \Majorana~ approach in the background reduction techniques through the selection of SSEs, \onbb-like events.  The~rejection of MSEs will be performed by applying anti-coincidence cuts and PSD methods, as illustrated in Sections~\ref{GERDA-HowTo} and~\ref{sec:majorana_bkg_suppression}. Figure~\ref{legend_bkg_projection}a shows the expected effectiveness of these cuts in terms of the reduction of the $\gamma$ rate due to the $^{238}$U, $^{232}$Th and $^{40}$K chains~\cite{Zsigmond:2020bfx}. Figure~\ref{legend_bkg_projection}b shows the projected contribution of the different sources to the overall total background. A~total background of 1~$\times~10^{-4}$~\ctsper is anticipated~\cite{Edzards:2020qdd}.  

\begin{figure}[H]
    \includegraphics[scale=0.5]{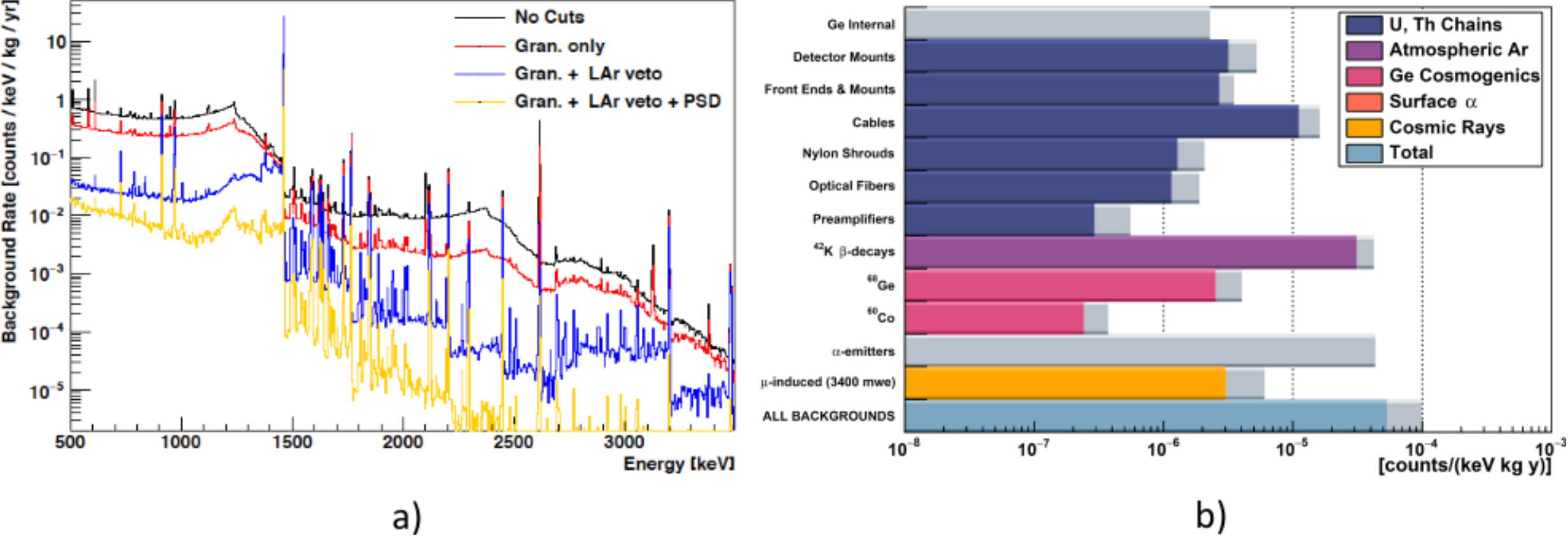}
    \caption{(\textbf{a}) Reduction of the $\gamma$ rate due to the $^{238}$U, $^{232}$Th and $^{40}$K chains. Figure published by J.~Phys. Conf. Ser., 2020~\cite{Zsigmond:2020bfx}. (\textbf{b})~Projected contribution of the different sources to the overall total background. Grey bars indicate 1$\sigma$ uncertainties of the background contributions due to screening measurements and Monte Carlo simulations. A~total background of 1~$\times~10^{-4}$~\ctsper~is anticipated. Figure published by J. Phys. Conf. Ser., 2020~\cite{Edzards:2020qdd}.}
    \label{legend_bkg_projection}
\end{figure}

\section{Conclusions}

In this paper, we reviewed the successful story of the experiments employing germanium semiconductor detectors in the search for the \onbb\ transition of \gesix. Starting from the pioneering work conducted by E.~Fiorini and collaborators in 1967, we followed the technology evolution from the original Ge(Li) diodes to the development of the modern HPGe detectors featuring a negligible intrinsic background contamination and impressive background rejection capabilities and energy resolution. Parallel to the development of the detectors, an~outstanding lowering of the background index by about eight orders of magnitude led to an increase in the sensitivity of the experiment by about six orders of magnitude in 50 years. Currently, the~\gerda\ experiment, implementing the use of bare Ge diodes immersed in LAr, succeeded at reaching the background-free regime, thus achieving the best sensitivity in the field with a lower exposure with respect to~competitors. 

The joint venture between \gerda\ and \majorana\  Collaborations led to establishing the \legend\ project, aiming at building a ton-scale experiment to fully span the inverted mass ordering region with a world-leading and timely competitive~program.    
\vspace{6pt}

\authorcontributions{Writing - original draft, Valerio D'Andrea, Natalia Di Marco, Carla Macolino, Michele Morella, Francesco Salamida; Writing - review \& editing, Valerio D'Andrea, Natalia Di Marco, Matthias Bernhard Junker, Matthias Laubenstein, Carla Macolino, Michele Morella, Francesco Salamida, Chiara Vignoli. All authors have read and agreed to the published version of the manuscript.%MDPI: For research articles with several authors, a short paragraph specifying their individual contributions must be provided. The following statements should be used ``Conceptualization, X.X. and Y.Y.; methodology, X.X.; software, X.X.; validation, X.X., Y.Y. and Z.Z.; formal analysis, X.X.; investigation, X.X.; resources, X.X.; data curation, X.X.; writing---original draft preparation, X.X.; writing---review and editing, X.X.; visualization, X.X.; supervision, X.X.; project administration, X.X.; funding acquisition, Y.Y. All authors have read and agreed to the published version of the manuscript.'', please turn to the  \href{http://img.mdpi.org/data/contributor-role-instruction.pdf}{CRediT taxonomy} for the term explanation. Authorship must be limited to those who have contributed substantially to the work~reported.
}

\funding{This research received no external funding %MDPI:Please add: ``This research received no external funding'' or ``This research was funded by NAME OF FUNDER grant number XXX.'' and  and ``The APC was funded by XXX''. Check carefully that the details given are accurate and use the standard spelling of funding agency names at \url{https://search.crossref.org/funding}, any errors may affect your future funding.
}

\institutionalreview{Not applicable %MDPI:In this section, please add the Institutional Review Board Statement and approval number for studies involving humans or animals. Please note that the Editorial Office might ask you for further information. Please add ``The study was conducted according to the guidelines of the Declaration of Helsinki, and approved by the Institutional Review Board (or Ethics Committee) of NAME OF INSTITUTE (protocol code XXX and date of approval).'' OR ``Ethical review and approval were waived for this study, due to REASON (please provide a detailed justification).'' OR ``Not applicable'' for studies not involving humans or animals. You might also choose to exclude this statement if the study did not involve humans or animals.
}

\informedconsent{Not applicable %MDPI:Any research article describing a study involving humans should contain this statement. Please add ``Informed consent was obtained from all subjects involved in the study.'' OR ``Patient consent was waived due to REASON (please provide a detailed justification).'' OR ``Not applicable'' for studies not involving humans. You might also choose to exclude this statement if the study did not involve humans.

}

\dataavailability{Data sharing not applicable %MDPI:In this section, please provide details regarding where data supporting reported results can be found, including links to publicly archived datasets analyzed or generated during the study. Please refer to suggested Data Availability Statements in section ``MDPI Research Data Policies'' at \url{https://www.mdpi.com/ethics}. You might choose to exclude this statement if the study did not report any data.
} 

\acknowledgments{The authors would like to thank Riccardo Brugnera and Steven Ray Elliott for their suggestions and comments. %MDPI:In this section you can acknowledge any support given which is not covered by the author contribution or funding sections. This may include administrative and technical support, or donations in kind (e.g., materials used for experiments).
}

\conflictsofinterest{The authors declare no conflict of interest. %MDPI:Declare conflicts of interest or state ``The authors declare no conflict of interest.'' Authors must identify and declare any personal circumstances or interest that may be perceived as inappropriately influencing the representation or interpretation of reported research results. Any role of the funders in the design of the study; in the collection, analyses or interpretation of data; in the writing of the manuscript, or in the decision to publish the results must be declared in this section. If there is no role, please state ``The funders had no role in the design of the study; in the collection, analyses, or interpretation of data; in the writing of the manuscript, or in the decision to publish the~results''.
}

\end{paracol}
\reftitle{References}
%\bibliographystyle{unsrt}
%\bibliography{biblio}

\end{document}